\documentclass[reprint,aps,prb,showpacs,amsmath,amssymb,superscriptaddress,nofootinbib,twocolumn]{revtex4-1}

\usepackage{color}
\usepackage{bm}
\usepackage{units}

\usepackage[caption=false]{subfig} 
\usepackage{bbold}
\usepackage{tikz}
\usepackage{graphicx}
\usepackage{bm}
\usetikzlibrary{calc}
\usetikzlibrary{arrows}
\usepackage{hyperref}
\usepackage{comment}
\DeclareSymbolFont{matha}{OML}{txmi}{m}{it}
\DeclareMathSymbol{\varv}{\mathord}{matha}{118}
\DeclareMathSymbol{\varu}{\mathord}{matha}{117}
\DeclareMathSymbol{\varw}{\mathord}{matha}{119}

\usepackage{bbold}
\usepackage{braket}
\usepackage{float}
\usepackage{xcolor}
\usepackage{changepage}
\usepackage{amsmath}

\DeclareMathOperator{\Tr}{Tr}

\usepackage{hyperref}
\usepackage{booktabs}
\makeatletter
\newcommand*{\rom}[1]{\expandafter\@slowromancap\romannumeral #1@}
\makeatother
\newcommand{\red}[1]{\textcolor{black}{#1}}

\begin{document}

\title{Shape dependence of entanglement negativity and mutual information in quantum Hall and critical systems}

\author{Chia-Chuan Liu}
\affiliation{Département de Physique, Université de Montréal, Montréal, Qc, H3C 3J7, Canada}

	\author{Juliette Geoffrion}
\affiliation{Département de Physique, Université de Montréal, Montréal, Qc, H3C 3J7, Canada}

\author{William Witczak-Krempa}
\affiliation{Département de Physique, Université de Montréal, Montréal, Qc, H3C 3J7, Canada}
\affiliation{
 Institut Courtois, Université de Montréal, Montréal, Qc, H2V 0B3, Canada
}
\affiliation{
 Centre de Recherches Mathématiques, Université de Montréal, Montréal, QC, HC3 3J7, Canada}

\begin{abstract}
We study two entanglement measures in a large family of isotropic many-body states including incompressible quantum Hall liquids and quantum critical systems: the logarithmic negativity (LN), and mutual information (MI). For pure states, obtained for example from a bipartition at zero temperature, these provide distinct characterizations of the entanglement present between two spatial subregions, while for mixed states (such as at finite temperature) only the LN remains a good entanglement measure. Our focus is on regions that have corners, either adjacent or tip-touching. We first obtain general non-perturbative properties regarding the geometrical dependence of the LN and MI. A close similarity is observed with mutual charge fluctuations, where super-universal angle dependence holds allowing for explicit checks. For the MI, we make stronger statements due to strong subadditivity. We also give ramifications of our general analysis to conformal field theories (CFTs) in two spatial dimensions.
We then explicitly verify these properties with integer quantum Hall states. To do so we develop two independent approaches to obtain the fermionic LN, which takes into account Fermi statistics: an overlap-matrix method, and a real-space lattice discretization. At finite temperature, we find a rapid decrease of the LN well inside the cyclotron gap at integer fillings. We further show that the LN decays faster compared to the MI at high temperatures.

\end{abstract}

\maketitle

\tableofcontents

\section{Introduction}\label{sec:intro}
The study of how entanglement is organized in complex many-body systems has led to new insights into quantum matter, ranging from the identification of topological states\cite{Kitaev2,Levin,HuiLi} to providing signatures for many-body localized systems.\cite{Prosen,Bardarson,Geraedts} Numerous approaches are based on the reduced density matrix of a subregion, and more particularly on the von Neumann entanglement entropy (EE). The EE quantifies the amount of quantum entanglement between two subregions on a bipartite geometry for pure states, such as groundstates. However, the EE does not correctly measure entanglement for mixed states, for instance thermal states, since it also contains classical correlations. Indeed, the EE, which usually obeys a boundary law in groundstates, reduces to the classical entropy at sufficiently high temperatures with its volume-law scaling. The mutual information (MI) between non-overlapping subregions $A_1$ and $A_2$ eliminates the volume law, but can still be ``polluted'' by classical correlations between the subregions. A quantity similar to the MI but with the advantage of only capturing entanglement, even for mixed states, is the logarithmic negativity (LN).\cite{zyczkowski98,Vidal, Plenio} It is obtained from a transposition of the density matrix $\rho_{A_1A_2}$ \red{with respect to one of the subsystems}, say $A_1$. In operational terms, it serves as an upper bound for distillable entanglement\cite{Vidal}, i.e.\ the number of Bell pairs one can extract from multiple copies of the state $\rho_{A_1 A_2}$. The LN has been studied, among others, for topological phases\cite{Castelnovo,Hart,Yirun,Liu_2022}, and quantum critical systems.\cite{Nobili,Calabrese_2013,Tonni,Lu,Shapourian_2019}

In this work, we study the LN in a large class of \red{quantum many-body} states, and geometries. We compare our findings with the MI, as well as a simpler quantity, mutual fluctuations \red{or uncertainty: a MI-analog built with the uncertainty of a local observable (such as charge) in a subsystem}. In particular, we focus on isotropic states, such as incompressible quantum Hall \red{many-body} states, or quantum critical states including conformal field theories (CFTs).  We briefly discuss \red{the case of} separated subregions, \red{illustrated in Fig.~\ref{fig:separated},} and then move on to various geometries with two corners that touch, either via an edge or \red{a} vertex. By using general considerations, such as the presence of a boundary law for adjacent subregions or the strong subadditivity (SSA) for the EE, we obtain numerous non-perturbative results for the angle dependence \red{of the LN, MI, and mutual fluctuations. We note that these strong results for the LN hold for both the bosonic and fermionic definitions of the LN.} 
We then verify our findings using integer quantum Hall (IQH) states at various fillings and temperatures. In doing so, we develop two distinct methods \red{to compute the fermionic LN in Hall states}: a momentum overlap-matrix method, as well as \red{a} real-space discretization method. The former is very accurate at low temperatures, whereas the latter is useful at finite temperatures. Interestingly, at finite temperature, we find a marked reduction of the LN well below the cyclotron gap, which is proportional to the applied magnetic field. 

The rest of the paper is organized as follows. 
In Sec.~\ref{sec:notion}, we introduce the notion of the partial transpose by which the LN is defined. 
In Sec.~\ref{sec:general}, we obtain the non-perturbative results regarding the geometrical dependence of the LN (both bosonic and fermionic), MI and mutual fluctuations. We explain the methodology of the numerical calculations for IQH states in Sec.~\ref{sec:method}. Sec.~\ref{sec:tripartieZeroT} shows the corner dependences of the LN and the MI on various tripartite geometries and fillings at zero temperature. Finally, in Sec.~\ref{sec:LNfiniteT}, we study the temperature dependence of the LN and MI.

\section{Logarithmic negativity}\label{sec:notion}

\subsection{Bosonic and fermionic partial transpose}
Let $\rho$ be the density matrix of a quantum system defined on a region $AB$. The reduced density matrix $\rho_A$  defined on  subsystem $A$ is $\rho_A=\Tr_B{\rho}$. The R\'enyi entropy of index $n>0$ for $A$ is defined as $S_n(A) = \frac{1}{1 - n} \log \Tr(\rho_A^n)$. The EE, $S(A) = - \Tr \rho_A \log \rho_A$, follows from the limit $n \rightarrow{}1$. Suppose we further divide subregion $A$ into two subregions $A_1$ and  $A_2$ so that $A=A_1 A_2$. The reduced density matrix $\rho_A$ can be expressed as
\begin{equation}
\rho_A=\sum_{ijkl}\langle e^{(1)}_i, e^{(2)}_j\vert\rho_A\vert e^{(1)}_k, e^{(2)}_l\rangle\vert  e^{(1)}_i, e^{(2)}_j\rangle \langle  e^{(1)}_k, e^{(2)}_l \vert
\end{equation}
where $\vert e^{(1)}_i\rangle$ and $\vert e^{(2)}_j\rangle$ denote orthonormal bases in the Hilbert spaces $\mathcal{H}_1$ and $\mathcal{H}_2$ corresponding to $A_1$ and $A_2$, respectively. The bosonic partial transpose (PT) with respect to $A_1$ is
\begin{equation}
(\vert e^{(1)}_i, e^{(2)}_j\rangle\langle e^{(1)}_k ,e^{(2)}_l\vert)^{T_1}=\vert e^{(1)}_k, e^{(2)}_j\rangle\langle e^{(1)}_i ,e^{(2)}_l\vert
\end{equation}
and the reduced density matrix $\rho^{T_1}_A$ after the bosonic PT becomes   
\begin{equation}\label{eq:BPT}
\rho^{T_1}_A=\sum_{ijkl}\langle e^{(1)}_k, e^{(2)}_j\vert\rho_A\vert e^{(1)}_i, e^{(2)}_l\rangle\vert  e^{(1)}_i, e^{(2)}_j\rangle \langle  e^{(1)}_k, e^{(2)}_l \vert.
\end{equation}
The LN $\mathcal{E}_b$ defined via the bosonic PT above is \cite{Vidal, Plenio}
\begin{equation}\label{eq:LNB}
\mathcal{E}_b=\log\Tr{\sqrt{\rho^{T_1}_A \left(\rho^{T_1}_A\right)^{\dagger}}}.
\end{equation}
It does not depend on whether we perform the partial transpose on subregion $A_1$ or $A_2$. 

If the reduced density matrix $\rho_A$ is separable, including mixed states, $\mathcal{E}_b$ vanishes. \cite{Peres, Horodecki} Such a separability criterion, known as the positive partial transpose (PPT) criterion, implies that a non-vanishing $\mathcal{E}_b$ is a 
sufficient condition for the presence of quantum entanglement. For pure states, $\mathcal{E}_b$ equals the R\'enyi entropy at index one-half, 
$S_{1/2}(A_1) = 2\log \Tr\sqrt{\rho_{A_1}}=S_{1/2}(A_2)$.\cite{Vidal} However, this relation ceases to hold for mixed states. Unlike the R\'enyi entropy and the EE, $\mathcal{E}_b$ (\ref{eq:LNB}) is an entanglement monotone \red{for general states, meaning} that it does not increase under local operations and classical communications (LOCC).\cite{Vidal,Plenio} Therefore, it can measure quantum entanglement even if the density matrix is mixed. 

However, for fermionic systems, the LN defined via the bosonic PT (\ref{eq:BPT}) $\mathcal{E}_b$ ignores sign changes that appear due to exchanging fermions which leads to certain limitations. For one, for a gaussian density matrix, the partial transpose leads to a non-gaussian density matrix\cite{Eisler}, which makes evaluating the LN a difficult task.\cite{coser} Furthermore, in some cases, applying the bosonic PT to fermionic systems underestimates the true degree of entanglement since it allows operations that violate fermionic number conservation to reduce the entanglement.\cite{Shapourian1} To remedy these limitations, one can define a fermionic version of the LN, that we denote $\mathcal{E}$, with the so-called fermionic PT. \cite{Shapourian1} Consider the element of a density matrix in the coherent state basis $\vert \lbrace  \xi _j\rbrace\rangle=e^{-\sum_j{ \xi _jf^{\dagger}_j}}\vert 0\rangle$, where $f^{\dagger}_j$ is a fermionic creation operator and $\xi_j$ is a Grassman variable. The fermionic PT transforms elements in the following manner: 
\begin{equation}\label{eq:FPT}
\begin{aligned}
&U_{A_1}\left(\vert \lbrace  \xi _j\rbrace_{j\in A_1}, \lbrace \xi _j\rbrace_{j\in A_2} \rangle \langle \lbrace \overline{\chi}_j\rbrace_{j\in A_1}, \lbrace\overline{\chi}_j\rbrace_{j\in A_2} \vert \right)^{\mathcal T_1}U^{\dagger}_{A_1}\\
&=\vert  \lbrace i\overline{\chi}_j\rbrace_{j\in A_1}, \lbrace  \xi _j\rbrace_{j\in A_2} \rangle \langle  \lbrace i
 \xi _j\rbrace_{j\in A_1},\lbrace \overline{\chi}_j\rbrace_{j\in A_2} \vert,\\
\end{aligned}
\end{equation}
where  $U_{A_1}$ is a unitary operator on subsystem $A_1$ related to the time-reversal operator. In the rest of the paper, we focus on $\mathcal{E}$ defined by the fermionic PT, Eq.~(\ref{eq:FPT}). It has been proven that in this case, the LN $\mathcal{E}$ is still an entanglement monotone and satisfies the separability criterion.\cite{Shapourian3} For any bipartite pure state, the bosonic and fermionic LN are equal, that is,  $\mathcal{E}(A,B)=\mathcal{E}_b(A,B)=S_{1/2}(A)=S_{1/2}(B)$\cite{footnote1}. However, for mixed states on $A B$, the bosonic and fermionic LN generally differ, with the fermionic LN acting as an upper bound to the bosonic one in the gaussian case. \cite{Wang, Zimboras}


Consider the fermionic PT (\ref{eq:FPT}) in the Majorana basis. Let $\lbrace m_1,m_2,...,m_{2k} \rbrace$ and $\lbrace n_1,n_2,...,n_{2l} \rbrace$ denote the indices of Majorana operators $a_x$ belonging to the subsystems $A_1$ and $A_2$ respectively, and introduce the notation $a^0_{x}=1$ and $a^{1}_x=a_x$. The density matrix $\rho_A$ for a fermionic state on $A=A_1 A_2$ can be expressed in term of Majorana operators as
\begin{equation}\label{eq:Dnmaj}
\rho_A=\sum_{\substack{\underline{\kappa},\underline{\tau},\\ \vert \kappa\vert+\vert\tau\vert=\text{even}}}w_{\underline{k},\underline{\tau}}a^{\kappa_1}_{m_1}\dots a^{\kappa_{2k}}_{m_{2k}} a^{\tau_1}_{n_1}\dots a^{\tau_{2l}}_{n_{2l}} 
\end{equation}
where $\underline{\kappa}=(\kappa_1,\dots,\kappa_{2k})$ and $\underline{\tau}=(\tau_1,\dots,\tau_{2l})$ in the summation run over all bit-strings of length $2k$ and $2l$, respectively, and $\vert\kappa\vert=\sum_{i}\vert\kappa_i\vert$. Note that since physical fermionic density operators must commute 
with the total fermion-parity operator, one has $w_{\underline{\kappa},\underline{\tau}}=0$ when $\sum_i\kappa_i+\sum_{j}\tau_j$ is odd. Based on the expression (\ref{eq:Dnmaj}), the density matrix $\rho_A$ transforms under the fermionic PT (\ref{eq:FPT})  on $A_1$  as
\begin{equation}\label{eq:PRDnmaj}
\rho^{\mathcal T_1}_A=\sum_{\substack{\underline{\kappa},\underline{\tau},\\\vert \kappa\vert+\vert\tau\vert=\text{even}}}i^{\vert\kappa\vert}w_{\underline{k},\underline{\tau}}a^{\kappa_1}_{m_1}\dots a^{\kappa_{2k}}_{m_{2k}} a^{\tau_1}_{n_1}\dots a^{\tau_{2l}}_{n_{2l}}. 
\end{equation}

For later convenience, we introduce  another normalized composite density operator $\rho_{\times}$:
\begin{equation}\label{eq:composeden}
\rho_{\times}=\frac{\rho^{\mathcal T_1}_{A}\left(\rho^{\mathcal T_1}_{A}\right)^{\dagger}}{Z_{\times}}
\end{equation}
where $Z_{\times}=\Tr{(\rho^{\mathcal T_1}_{A}(\rho^{\mathcal T_1}_{A})^{\dagger})}=\Tr{\rho^2_A}$, so that the LN defined via the fermionic PT (\ref{eq:FPT}) can be expressed as \cite{Zimboras} 
\begin{equation}\label{eq:negPR2}
\begin{aligned}
&\mathcal{E}=\log\Tr{\sqrt{\rho^{\mathcal T_1}_A\left(\rho^{\mathcal T_1}_A\right)^{\dagger}}} \!=\log\Tr{\rho^{\frac{1}{2}}_{\times}}+\frac{1}{2}\log{\Tr{\rho^2_A}}.
\end{aligned}
\end{equation}

\red{\section{Non-perturbative shape dependence}}
\label{sec:general}
We will describe salient features of the LN, $\mathcal E(A_1,A_2)$ \red{taken here to denote both the bosonic and fermionic definitions}, and the MI, 
\begin{align}
    I(A_1,A_2)=S(A_1)+S(A_2)-S(A_1 A_2)
\end{align}
in a large class of isotropic states, including incompressible quantum Hall groundstates. Our general results apply equally well to the fermionic definition of the LN $\mathcal E$, Eq.~(\ref{eq:negPR2}), and to the bosonic one $\mathcal E_b$, Eq.~(\ref{eq:LNB}). We compare the LN and MI with a simpler quantity: the mutual  fluctuations or variance, \red{which is defined analogously to the MI:} 
\begin{align}
    \mathcal I(A_1,A_2)=\mathcal F(A_1)+\mathcal F(A_2)-\mathcal F(A_1 A_2)
\end{align}
where 
\begin{align}\label{eq:bf} \red{
    \mathcal F(A) = (\Delta Q_A)^2=  \langle Q_A^2\rangle - \langle Q_A\rangle^2 }
\end{align} 
gives the bipartite fluctuations or variance, \red{namely quantum uncertainty squared,} of a local observable $\varrho(\vec r\,)$ in a subregion $A$, \red{$Q_{A}=\int_{A}\! d\vec r \varrho$. The expectation values in \eqref{eq:bf} are taken
with respect to the reduction of a density matrix $\rho$ defined on $AB$, $\langle O_A\rangle=\Tr(O_A\rho_A)$, where the reduced density matrix is $\rho_A=\Tr_{B}\rho$}. 
\red{In contrast to the LN and MI, $\mathcal I(A_1,A_2)$ is not necessarily positive so that it can increase under LOCC. Negative mutual fluctuations can occur even in the case of a conserved charge 
in a pure state $|\psi\rangle_{A_1 A_2 B}$. For instance, let us consider the following 4-qubit cat state:
\begin{align}
    |\psi\rangle=(|\!++--\rangle+|\!--++\rangle)/\sqrt2, 
\end{align}
where $A_1$ is the first qubit, while $A_2$ is the second one; $B$ is composed of the third and fourth qubits. It is an eigenstate of the total spin along $z$, $Q=\sum_{i=1}^4 Z_i$, where $Z_i$ is the Pauli-$Z$ operator for the $i$th qubit ($Z_i|\pm\rangle=\pm|\pm\rangle_i$). One readily obtains the negative mutual fluctuations $\mathcal I(A_1,A_2)=-2$. For the family of isotropic density matrices under consideration, we will nevertheless see that mutual fluctuations share numerous properties with both the LN and MI.} 


\begin{figure}
\centering
\includegraphics[scale=0.3]{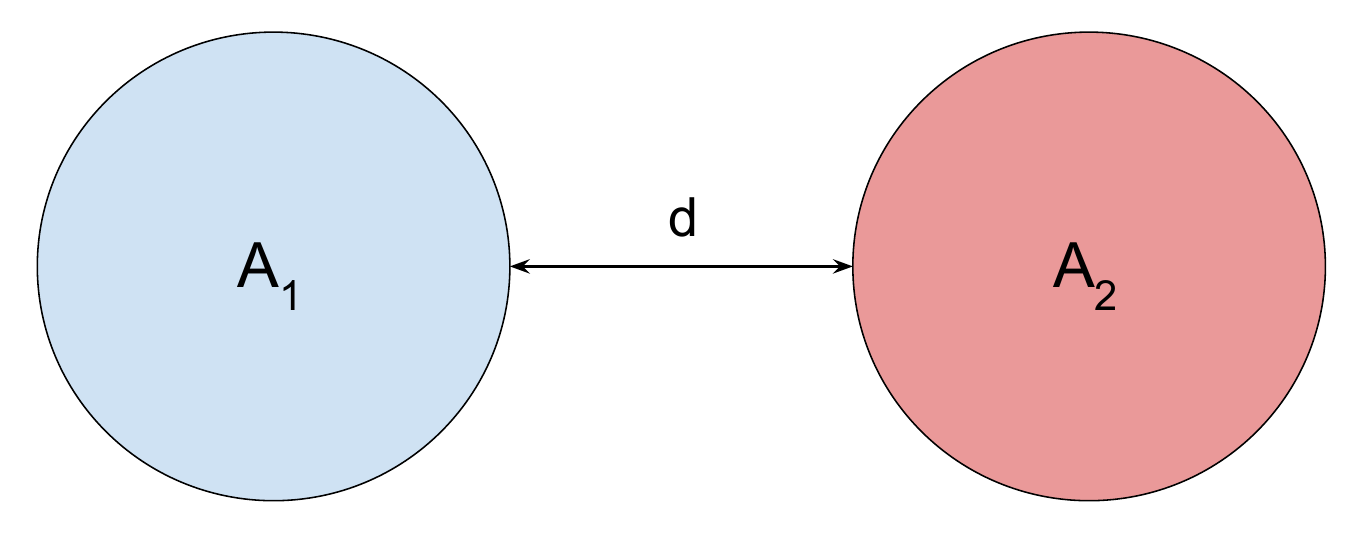}
\caption{Subregions $A_1$ and $A_2$ are separated by a distance $d$. The complement of $A_1 A_2$ is $B$ (white region). \label{fig:separated} } 
\end{figure}    
\subsection{Separated regions} 
The \red{local properties of the states under consideration dictate that the} LN will decrease when the separation between $A_1$ and $A_2$ increases. In scale invariant states, like quantum critical ground states (an example being CFTs), this will necessarily occur as a power law \red{at long distances}. For quantum Hall groundstates, the decay will be exponential due to the gap.  Let $d$  be the scale that determines the separation between subregions $A_1$ and $A_2$, as depicted in Fig.~\ref{fig:separated}. We then expect $\mathcal E= \mathcal O(\exp(- \zeta d))$, where $\zeta$ is a positive coefficient inversely proportional to the gap. For IQH groundstates at any  integer filling $\nu$, one can go a step further. For the mutual fluctuations of charge, it is easy to see that they decay with a Gaussian envelope $\exp\left(-\tfrac12(d/l_B)^2\right)$, where we have reinstated the magnetic length. This occurs due to the Gaussian decay of the electronic Green's function $C_{\mathbf{r},\mathbf{r}'}$. Numerically, we observe that the LN also has such a Gaussian envelope at large separations $d$, thus decaying much faster than the naive guess $\exp(- \zeta d)$. It is indeed easy to see that when the Green's function vanishes between regions $A_1$ and $A_2$, $C_{12}\simeq 0$, the LN vanishes (see Appendix \ref{appendix:LNnocorrelation}). At large but finite separations, the Green's function between regions 1 and 2 $C_{12}$ has Gaussian-suppressed matrix elements, and these will lead to a \red{small} Gaussian-suppressed LN. 

We now turn to \red{simple and important} geometries, \red{namely those} with sharp corners, and examine the resulting angle-dependence. We obtain numerous non-perturbative results that reveal a similar structure among \red{the LN, MI and mutual fluctuations}. 

\subsection{Single corner}
\begin{figure} 
\centering
\includegraphics[scale=0.36]{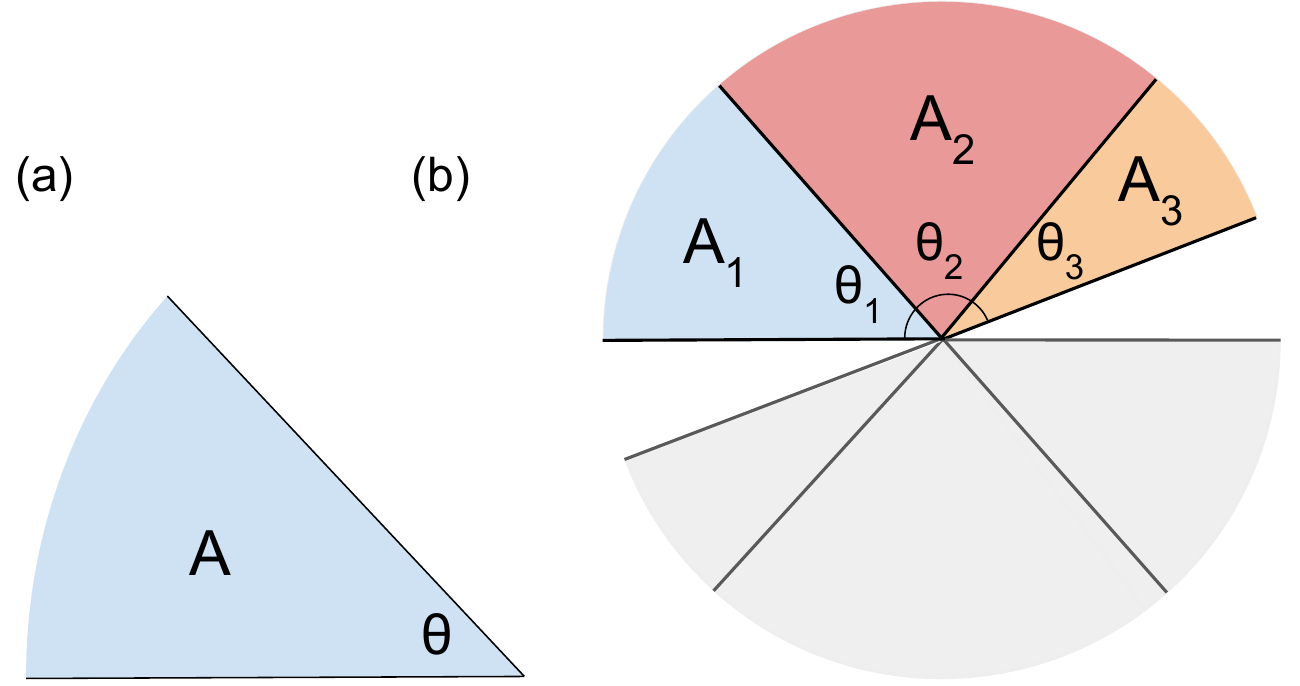}
\caption{\label{fig:single-corner} Partitions with corners. (a) Subregion  $A$ with a single corner of angle $\theta$; the complement $B$ is the complementary corner of angle $2\pi-\theta$. 
(b) The regions $A_{1,2,3}$ are used to show that the corner function for the EE, $a(\theta)$, is convex. 
The inversion of $A_1$ about its apex is shown in grey; the union of $A_1$ and its inverse \red{image form} an hourglass with tip-touching corners. Similarly for $A_2$ and $A_3$.} 
\end{figure}    
Let us first examine the \red{simplest} case of a subregion with a single corner of angle $\theta$, as shown in in Fig.~\ref{fig:single-corner}(a), before turning to the case of two non-complementary corners in the next subsection.
The EE \red{of a state with finite correlations, such as the groundstate of a gapped Hamiltonian,} can be expanded as follows in the large $|A|$ limit:
\begin{align} \label{eq:S}
	S(A) = \alpha_0 |A| + c_1 |\partial A| - a(\theta) - \gamma + \cdots
\end{align} 
where \red{$\alpha_0>0$ is the volume-law coefficient, $c_1$ the area-law coefficient}, $a(\theta)$ is the corner contribution, which vanishes at $\theta=\pi$ \red{when the corner disappears}. The last term  $\gamma$ denotes a topological contribution, while the dots correspond to subleading terms that vanish as \red{the size of the subregion grows}. An analogous expansion holds for the R\'enyi entropies, and bipartite fluctuations $\mathcal F$, \red{Eq.}~\eqref{eq:bf}. The corner term has been extensively studied including in quantum critical states (especially CFTs),\cite{Casini_2009,Hirata1,Kallin_2014,Stoudenmire_2014,Kovacs_2012,Bueno2,Bueno1,Helmes, Whitsitt,WWK19}  and topological phases.\cite{Rodriguez2,Sirois,Estienne}
We can use SSA to show that the EE single-corner term is convex 
\begin{align} \label{eq:a-conv}
	a''(\theta)\geq 0, 
\end{align}
which was numerically observed in Ref.~\onlinecite{Sirois} for IQH groundstates at fillings $\nu = 1,2$, as well as for an excited state at unit filling. The argument is adapted and generalized from the one given for the groundstates CFTs in (2+1) spacetime dimensions\cite{Hirata1} \red{but does not require Lorentz invariance or vanishing temperature.} Let us consider three adjacent corners $A_1,A_2,A_3$ of angles  $\theta_1,\theta_2,\theta_3$, respectively, as shown in Fig.~\ref{fig:single-corner}(b). SSA can be formulated as the following inequality: $S(A_1+A_2+A_3) + S(A_2)\leq S(A_1+A_2)+S(A_2+A_3)$.
First, the volume and boundary law terms cancel.
Second, since all the combinations of subregions appearing in the inequality have the same topology, the topological terms also cancel.  One is then left with the following inequality for the corner terms:
\begin{align} \label{eq:cornerSSA}
	a(\theta_1+\theta_2+\theta_3)-a(\theta_1+\theta_2)  \geq   a(\theta_2+\theta_3)- a(\theta_2).
\end{align}
Taking first $\theta_3\to 0$, leads to $a'(\theta_1+\theta_2)\geq a'(\theta_2)$. Finally, taking $\theta_1\to 0$ leads to convexity: $a''(\theta)\geq 0$ for all angles $0\leq \theta\leq 2\pi$.

Let us now momentarily restrict ourselves to states that are pure on the entire space $A B$. This leads to the complementarity relation $S(A)=S(B)$, which implies $a(2\pi-\theta)=a(\theta)$.
We can also show that $a(\theta)$ is strictly decreasing for angles less than $\pi$ by setting $\theta_1=2\pi-2\theta_2$ in (\ref{eq:cornerSSA}), with $0\leq \theta_2 \leq \pi$. 
Combining this with the complementarity, and taking $\theta_3\to 0$ yields $a'(\theta)\leq 0$ for $0\leq \theta \leq \pi$. Now, since $a(\pi)=0$, this shows that the corner function is non-negative $a(\theta)\geq 0$. Summarizing, for pure states,
\begin{align} \label{eq:a-pure}
a(\theta)\geq 0\, , \enspace a'(0\leq\theta\leq \pi)\leq 0 \,, \enspace a(\theta)=a(2\pi-\theta).
\end{align}

\red{Returning to general (mixed)} states, the corner function vanishes in the absence of a corner, $a(\pi)=0$. Since this limit is not singular, we can Taylor expand about $\pi$. For pure states, only even powers appear due to complementarity:
\begin{align}
	a(\theta)=\sigma\,(\theta-\pi)^2 + \sigma' \, (\theta-\pi)^4+\cdots, 
\end{align}
where $\sigma\geq 0$ due to convexity, Eq.~\eqref{eq:a-conv}. For general (mixed) states, 
odd powers cannot be ruled out from the current analysis. In that case, we nevertheless have $\sigma\geq 0$ since convexity \eqref{eq:a-conv} holds for general density matrices.  

In the opposite limit of small angles, $\theta\to 0$,  the EE $S(A)$ must be decreasing since the amount of entanglement or correlations is limited by \red{the number of} degrees of freedom in $A$, \red{which vanishes in the zero-angle limit}.
However, in the pie-shape geometry of Fig.~\ref{fig:single-corner}(a), the boundary law contribution is not changing as the angle is decreasing. The corner term must thus \emph{effectively} counteract it \red{to give a decreasing EE}: $a(\theta\to 0)\sim |\partial A|/\delta_{\rm eff}$, where $\delta_{\rm eff}$ is an effective short-distance cutoff proportional to $\theta\ll 1$: $\delta_{\rm eff}\sim \theta L$, where $|\partial A|= 2 L$ is the fixed perimeter of the subregion. \red{We emphasize that we are not saying the corner term exactly cancels the boundary law, since in the continuum limit this effective cutoff $\delta_{\rm eff}$ cannot be taken to be as small as the true UV cutoff.} We thus have \red{the following divergence for the corner term}
 \begin{align} \label{eq:a-sharp}
 	a(\theta\to 0) = \frac{\kappa}{\theta}
 \end{align}
 where \red{we introduced the quantity} $\kappa \geq 0$, a state-dependent coefficient. Precisely the same divergence will occur as $\theta\to2 \pi$ since in that limit the corner term also effectively acts to cancel the boundary law. \red{The result \eqref{eq:a-sharp} respects the convexity relation \eqref{eq:a-conv}, $a''(\theta)\geq 0$.} 
 The behaviour \eqref{eq:a-sharp} was numerically confirmed for IQH groundstates at fillings $\nu=1,2$ as well as for an excited state, 
 and the $\kappa$ coefficients \red{were} obtained.\cite{Sirois}  
A diverging corner contribution \red{in the small-angle limit} was also obtained for the 2nd R\'enyi entropy $S_2$ for a fractional quantum Hall state at filling $\nu=1/2$ for bosons.\cite{Estienne} 

In the case of bipartite fluctuations $\mathcal F(A)$, we have an expansion as in Eq.~(\ref{eq:S}) but the corner term possesses a \red{strikingly simple} super-universal angle dependence\cite{Estienne}
\begin{align} \label{eq:a-fluc}
 a_{\rm fluc}(\theta) = \alpha \left(1+ (\pi-\theta)\cot\theta \right)
\end{align}
where the state-dependent information is entirely encoded in the \red{sum-rule} coefficient $\alpha=-\int_0^\infty dr\, r^3 f(r)/2$, with \red{$f(r)=\langle\varrho(\bm r)\varrho(0)\rangle-\langle\varrho(\bm r)\rangle \langle\varrho(0)\rangle$ being the connected correlation function ($r=|\bm r|$).} We find
the  small-angle divergence in Eq.~(\ref{eq:a-sharp}), with $\kappa_{\rm fluc}=\pi\alpha$. \red{When the coefficient $\alpha$ is non-negative (as occurs in numerous states of physical interest,\cite{Estienne} see quantum Hall examples below), one can check that all properties given above for the EE are respected.}

\subsubsection{Mutual quantities} 
When the two regions share a boundary of length $L_{\rm shared}$, the LN, MI and fluctuations will scale with the length of the \red{shared} boundary. For the MI, this follows from the fact that volume law contributions are cancelled, leaving behind the boundary law contributions along the shared boundary. For example, for the LN in a pure state, when $A B$  is the entire system, we will get $\mathcal E=S_{1/2}(A)=S_{1/2}(B)=c_{1/2}L_{\rm shared}+\cdots$, which is dominated by the boundary law.
For the geometry where $A$ is a corner of angle $\theta$ and $B$ is the complementary corner, as depicted in Fig.~\ref{fig:single-corner}(a), the LN, MI and mutual fluctuations will scale as 
\begin{align}
	\mathcal E(A,B) &= c_{\mathcal E} L_{\rm shared} - a^{\mathcal E\!}(\theta), \\
	I(A,B) &= 2c_{1} L_{\rm shared} - a^{I\!}(\theta), \\
	\mathcal I(A,B) &= 2c_{\rm fluc} L_{\rm shared} - a^{\mathcal I\!}(\theta)
\end{align}
where we have omitted subleading terms. For the MI, we can express the corner term in terms of the one appearing in the EE: $a^{I\!}(\theta)=a(\theta)+a(2\pi-\theta)$; we have an analogous relation for mutual fluctuations $a^{\mathcal I}$. When the angle approaches zero, the same argument as above \red{(the corner contribution counteracts the area law)} yields a pole for all three quantities:
\begin{align}
    \mathcal E, I,\mathcal I\to \{\kappa^{\mathcal E},\kappa^I,\kappa^{\mathcal I} \} / \theta \;\mbox{  as  }\; \theta\to 0
\end{align}
\red{Since these are mutual measures, one is led to conclude that the same coefficient dictates the pole at $\theta=2\pi$.}

If the state is pure on $A B$, then the LN is given by the $1/2-$R\'enyi entropy
\begin{align}
    \mathcal E(A,B) = S_{1/2}(A)= c_{1/2} L_{\rm shared} - a_{1/2}(\theta)
\end{align}
so that $\kappa^{\mathcal E}=\kappa_{1/2}$. 

\begin{figure} 
\centering
\includegraphics[scale=0.33]{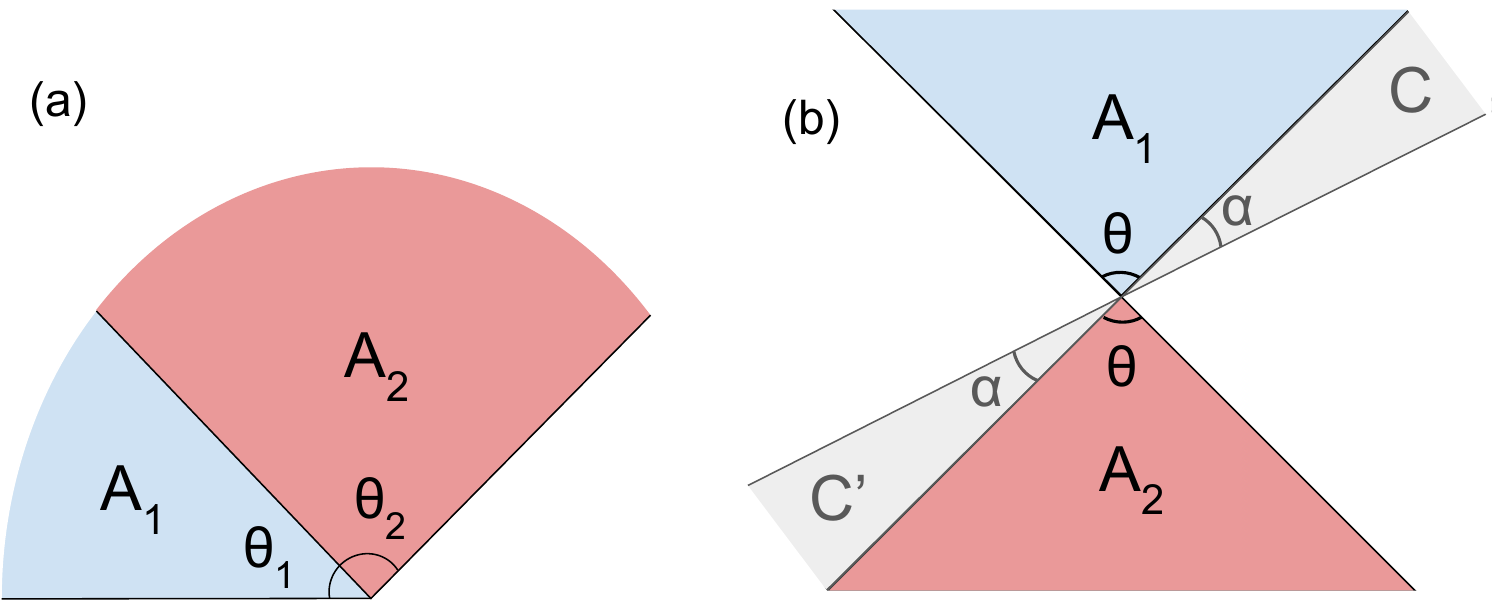}
\caption{Geometries where two corners, $A_1$ and $A_2$, touch. (a) The corners of angle $\theta_1$ and $\theta_2$ touch along an edge. (b) In this symmetric hourglass,
two corners have the same angle, and the geometry possesses an inversion symmetry about the apex. The grey regions $C,C'$ are used to show that the hourglass MI is monotonically increasing with $\theta$. 
\label{fig:two-corners} } 
\end{figure}  
\subsection{Adjacent corners}\label{sec:adj-general}
We now consider the case where two corners of angles $\theta_1$ and $\theta_2$, \red{not complementary $\theta_1+\theta_2<2\pi$,} are adjacent, as illustrated in Fig.~\ref{fig:two-corners}(a). 
There will be a new subleading corner term for mutual measures between $A_1$ and $A_2$:
\begin{align}
	\mathcal E &=c_{\mathcal E} L_{\rm shared} - b(\theta_1,\theta_2) + \cdots \\
	I &= 2c_1 L_{\rm shared} - b^{I\!}(\theta_1,\theta_2) + \cdots \\
	\mathcal I &= 2c_{\rm fluc} L_{\rm shared} - b^{\mathcal I}(\theta_1,\theta_2) +\cdots
\end{align} 
where for the MI
\begin{align} \label{eq:MI-adj}
	b^{I\!}(\theta_1,\theta_2) &= a(\theta_1)+a(\theta_2) - a(\theta_1+\theta_2)  
\end{align} 
with the analogous equation for $b^\mathcal I$ in terms of $a_{\rm fluc}(\theta)$, Eq.~(\ref{eq:a-fluc}).
When the shared boundary possesses more than one such corner, a sum over the corners appears \red{owing to locality}, $\sum_i b(\theta_1^{(i)}, \theta_2^{(i)})$ for the LN, similarly for the MI and mutual fluctuations. 

We shall now describe some limits in order to better understand the adjacent corner contribution. 
When one of the angles, say $\theta_1$, approaches zero, the entire LN $\mathcal E$ should decrease since $A_1$ is becoming vanishingly small, leading to fewer degrees of freedom, and so the amount of entanglement and correlations between $A_1$ and $A_2$ should correspondingly decrease. In fact, since the boundary law is not changing, the corner term $b$ has to grow in a way to effectively \red{counteract} the boundary law contribution. We thus expect $b(\theta_1,\theta_2)\sim L_{\rm shared}/\delta_{\rm eff}$, where $\delta_{\rm eff}$ is an effective cutoff that encodes the width of the shrinking region $A_1$. We can estimate $\delta_{\rm eff} \sim L_{\rm shared}\theta_1$, which leads to $1/\theta_1$ divergence: 
\begin{align} \label{eq:LN-pole}
 b(\theta_1\to 0,\theta_2)=k_{\rm adj} / \theta_1
\end{align}
where $k_{\rm adj}>0 $ is a state-dependent coefficient that is independent of $\theta_2$ since the latter remains finite, $\theta_2/\theta_1\to \infty$. Similarly, for the MI and mutual fluctuations we \red{can explicitly verify the above divergence via \eqref{eq:MI-adj}}: 
\begin{align} \label{eq:MI-pole}
 b^I(\theta_1\to 0,\theta_2)=\frac{\kappa}{ \theta_1}\,, \quad  b^{\mathcal I}(\theta_1\to 0,\theta_2)=\frac{\kappa_{\rm fluc}}{ \theta_1}
\end{align}
where we have used that $\kappa_{\rm adj} =\kappa$ and $\kappa^{\rm adj}_{\rm fluc}=\kappa_{\rm fluc}$ are the single-corner coefficients in the small angle limit. 

Furthermore, when the two adjacent angles add to $\pi$, $b(\theta,\pi-\theta)$ will be even about $\pi/2$ due to the symmetry exchanging the two subregions $1\leftrightarrow 2$.  
Owing to the small-angle divergences at $\theta=0$ and $\pi$, we thus expect a minimum at $\theta=\pi/2$. The entire LN is indeed expected to be maximal when both regions have the same size compared to the case where one of the regions is depleted at the expense of the other (thus possessing less degrees of freedom that can become entangled). Deviating from $\theta=\pi/2$, should thus lead to an increase of $b$, and a corresponding decrease of the LN (same for $I$ and $\mathcal I$). 
This point being non-singular, we can thus expand in even powers about $\pi/2$ :
\begin{align} \label{eq:adj-smooth}
 b(\theta,\pi-\theta) = b\!\left(\frac{\pi}{2},\frac{\pi}{2}\right) +  \sigma_{\mathcal E} (\theta-\frac{\pi}{2})^2+ \sigma_{\mathcal E}' (\theta-\frac{\pi}{2})^4+\cdots
\end{align}
with state-dependent coefficients $\sigma_{\mathcal E}$, $\sigma_{\mathcal E}'$. The minimum requirement at $\pi/2$ leads to a positivity constraint, $\sigma_{\mathcal E}\geq 0$. 
\red{Using the super-universal relation, Eq.~\eqref{eq:a-fluc}}, the above properties can be shown to hold explicitly \red{and rigorously} for mutual fluctuations $\mathcal I$. For the MI, we can \red{prove} convexity on general grounds for all angles:
\begin{align}
 \partial_\theta^2 b^I(\theta,\pi-\theta) = \partial_\theta^2(a(\theta)+a(\pi-\theta)) \geq 0
\end{align}
which follows from convexity of the single-corner term $a''(\theta)\geq 0$, Eq.~(\ref{eq:a-conv}). When combining the convexity with the reflection symmetry about $\pi/2$ and the divergences at $\theta=0,\pi$, we conclude that $b^I(\theta,\pi-\theta)$ indeed has a minimum at $\pi/2$. Moreover, this also means that $b^I(\theta,\pi-\theta)$ is decreasing on $[0,\pi/2]$:
\begin{align} \label{eq:b-decrease}
\partial_\theta b^I(\theta,\pi-\theta) \leq 0.
\end{align}
In the case where the state is pure on the entire space, the above inequality follows directly from Eq.~(\ref{eq:a-pure}).

\subsection{Hourglass} \label{sec:hg-general}
We now consider a geometry where the two corners are tip-touching, instead of adjacent, as shown Fig.~\ref{fig:two-corners}(b). 
This geometry has the advantage of removing the boundary law contribution. For simplicity, we shall consider the case of the symmetric hourglass. 
For the MI, we have
\begin{align} \label{eq:MI-hourglass}
I(\theta) &= a_\times(\theta) - 2a(\theta)
\end{align}
where $a_\times(\theta)$ is a new corner term associated with the EE an hourglass, whereas $a(\theta)$ is the usual corner coefficient of a single corner of angle $\theta$, Eq.~(\ref{eq:S}). 
The same structure arises for fluctuations $\mathcal I$.

As the angle $\theta$ approaches zero, we expect the LN and MI to vanish because most parts of region $A_1$ become very distant from $A_2$. For instance, in quantum Hall states, the parts of regions $A_1$ and $A_2$ that are within a magnetic length of the apex are shrinking to zero, which excludes sharing entanglement or correlations. By positivity of the LN and MI, we then conclude that $\mathcal E'(\theta), I'(\theta)\geq 0$ at small angles in isotropic states. 

In the opposite limit, $\theta\to \pi$, the boundary of $A_1$ becomes very close to that of $A_2$. This will mean that the LN and MI should start being dominated by an effective boundary law $\sim L_{\rm shared}/\delta_{\rm eff}$,  where $L_{\rm shared}$ is the length of the shared boundary at $\theta=\pi$. Since the regions are not touching, $\delta_{\rm eff}$ is not the short distance scale (or UV cutoff) as in the adjacent case, but rather a measure of separation between the two subregions. A simple geometric estimate gives $\delta_{\rm eff} /L_{\rm shared}\sim (\pi-\theta)$, which then gives rise to a pole at $\pi$:
\begin{align}
 \mathcal E(\theta\to\pi) &= \frac{k_\times }{\pi-\theta}  \label{eq:LN-pi}, \\
I(\theta\to\pi) &= \frac{\kappa_\times }{\pi-\theta}   \label{eq:MI-pi}
\end{align}
where $k_\times, \kappa_\times \geq 0$ are state-dependent coefficients. 

Furthermore, for pure states on the entire space $A_1A_2B$ we can deduce the \red{MI} coefficient $\kappa_\times$ from the single-corner function $a(\theta)$:
\begin{align} \label{eq:kappax}
\kappa_\times = 2\kappa.
\end{align}
The argument is the following. The corner function $a(\theta)$ vanishes in the $\theta\to\pi $ limit, so that $I(\theta)\approx a_\times(\theta)$. Now, if the density matrix on the entire space is pure, we have the complementarity relation $a_\times(\theta)=a_\times(\pi-\theta)$, which holds true for all $\theta$. When $\theta\to\pi$, the complementary hourglass function $a_\times(\pi-\theta)$ is evaluated for small angles. In that limit, the contributions from both halves of the hourglass decouple due to increasing spatial separation between degrees of freedom in $A_1$ and $A_2$, leading to $a_\times(\bar\theta)=2a(\bar\theta)=2\kappa/\bar\theta$, where $\bar\theta=\pi-\theta\ll 1$. We have thus arrived at the desired relation \eqref{eq:kappax}. 
We shall see that the relation (\ref{eq:kappax}) indeed holds for IQH states. It was also found to hold for certain \red{strongly coupled} large-$N$ supersymmetric CFTs described by the holographic AdS/CFT duality.\cite{Omidi} 

Knowing that the LN and MI both increase at small and large angles (the latter due to the divergence at $\pi$, Eq.~(\ref{eq:MI-pi})), we can inquire about what happens at intermediate angles. 
It is natural to expect both the LN and MI to increase monotonically for all $\theta$. 
\red{We can in fact prove it} for all angles for the MI by using SSA: 
\begin{align} \label{eq:MI-ssa}
    I(A,BC)\geq I(A,B)
\end{align}
where the subregions $A,B,C$ are embedded in the entire system. Consider the case $A=A_1$, $B=A_2$ and $C'$ is a pie-shaped region of angle $\alpha$ adjacent to $A_2$. We also introduce $C$, which is a pie-shaped region of angle $\alpha$ adjacent to $A_1$ but opposite to $C'$. This geometry is illustrated in 
Fig.~\ref{fig:two-corners}(b).
We thus have an enlarged hourglass with $I(A_1 C,A_2 C')= I(\theta+\alpha)$. By SSA, $I(A_1 C,A_2 C')\geq I(A_1,A_2)$, or $I(\theta+\alpha)\geq I(\theta)$. Taking the limit $\alpha\to 0$, gives 
\begin{align}
I'(\theta)\geq 0 
\end{align}
implying that the MI is monotonically increasing for all angles. Since the LN does not obey SSA in general, we \red{cannot employ the same proof as for the MI. However, we can invoke isotropy. As the angle $\theta$ increases, subregions 1 and 2 contain more local degrees of freedom with the same properties (isotropy), and these become closer to each other. 
The quantum entanglement between 1 and 2 captured by the LN must thus increase:
\begin{align} \label{eq:ln-increase}
    \mathcal E'(\theta)\geq 0
\end{align}
This reasoning can also be used for the MI, but in that case the correlations are both classical and quantum.
The result \eqref{eq:ln-increase} is in perfect agreement with the above asymptotics obtained for angles near $0$ and $\pi$}. Furthermore, in all cases studied below, we do observe a strict monotonic increase for both the MI and LN. 

We can also ask about the convexity of the MI, i.e.\ the sign of $I''(\theta)$. Let us begin with the hourglass corner term \red{of the EE} that appears in Eq.~(\ref{eq:MI-hourglass}), $a_\times(\theta)$. The argument proceeds as for the convexity of $a(\theta)$ but by replacing $A_1$ by the hourglass of angle $\theta_1$ formed by $A_1$ and its inverse image, and so on for $i=2,3$, as shown in Fig.~\ref{fig:single-corner}(b). One extra constraint is $\theta_1+\theta_2+\theta_3\leq \pi$. The boundary law and topological terms cancel, and we get the analog of (\ref{eq:cornerSSA}): 
$a_\times(\theta_1+\theta_2+\theta_3)-a_\times(\theta_1+\theta_2)  \geq   a_\times(\theta_2+\theta_3)- a_\times(\theta_2)$. Taking the same limits, $\theta_3\to 0$ then $\theta_1\to 0$, yields 
\begin{align} \label{eq:a-times-convex}
a_\times''(\theta) \geq 0.
\end{align}

Restricting our attention to density matrices that are pure on the entire space, we can obtain another relation by setting $\theta_1=\pi-2\theta_2$ with $\theta_2\leq \pi/2$: $a_\times(\pi - \theta_2+\theta_3)-a_\times(\pi - \theta_2)  \geq   a_\times(\theta_2+\theta_3)- a_\times(\theta_2)$. But the complementarity relation for the hourglass corner is $a_\times(\pi-\theta)=a_\times(\theta)$, so that $a_\times(\theta_2-\theta_3)\geq a_\times(\theta_2+\theta_3)$, i.e.\ 
$a_\times'(\theta)\leq 0$ for $0\leq \theta\leq \pi/2$. We now summarize the \red{additional} relations for the hourglass corner for density matrices that are pure on the entire space:
\begin{align} \label{eq:a-times-properties}
	a_\times(\pi-\theta)=a_\times(\theta)\,, \;\enspace a_\times'(\theta) \leq 0\;\; (0\leq\theta\leq \frac{\pi}{2}) 
\end{align}
Going back to the convexity of the MI for the hourglass geometry, we see that $I''(\theta)=a''_\times(\theta)-2a''(\theta)$, which means that we subtract a positive number from another positive number. 
In principle, the outcome could be negative so we cannot make a general statement for all angles at this point. 
If we look at the limit $\theta\approx \pi$, then $I''(\theta)\approx \partial_\theta^2[2\kappa /(\pi-\theta)]= 4\kappa/(\pi-\theta)^3$, which is positive since $\kappa\geq 0$. The convexity thus holds at sufficiently large angles. Owing to Eq.~(\ref{eq:LN-pi}), the LN has the same divergence near $\pi$, hence it is also convex for angles near $\pi$.
In the IQH groundstates studied in this work, we find that convexity holds for all angles. 

\subsection{Generalization to CFTs}

Let us now consider the groundstates of a family of strongly correlated gapless systems, CFTs. As above, our conclusions remain valid for both bosonic and fermionic definitions of the LN. 
First, we note that when $A$ has a corner, it is known that the R\'enyi entropy contains a corner term that is logarithmically divergent $S_n(A)= c_n |\partial A| - a_n(\theta)\log(|\partial A|/\delta)+\cdots$, where $\delta$ is a UV cutoff. The corner function is thus protected from UV details \red{by the logarithm, and is thus determined by CFT data}. In the case where $B$ is the complement of $A$, we have $\mathcal E(A,B)=S_{1/2}(A)$, and the LN will also possess a logarithmically divergent corner contribution with prefactor $a_{1/2}(\theta)$. In the case of the free scalar and Dirac fermion CFTs, exact results in the nearly-smooth limit $\theta\approx \pi$ are available due to a special boson-fermion duality,\cite{Bueno-twist} and at other angles from lattice numerics.\cite{Helmes,Tonni}
Fluctuations of a conserved charge will also have a logarithmically divergent corner term.\cite{Estienne, Herviou} These logarithmic enhancements will also appear when two corners meet, in particular for the adjacent and hourglass geometries, as we now discuss.

\subsubsection{Adjacent corners}
For adjacent corners,  we have an expression similar to what we found above, but with a logarithmic enhancement due to corners
\begin{align}
	\mathcal E =c_{\mathcal E} L_{\rm shared} - b(\theta_1,\theta_2) \log(L_{\rm shared}/\delta) + \cdots 
\end{align}
This was verified explicitly for the bosonic LN using the free scalar CFT realized as lattice of coupled quantum harmonic oscillators.\cite{Nobili} When one of the angles approaches zero, the LN and MI will also possess a pole, as in Eq.~(\ref{eq:LN-pole}) and Eq.~(\ref{eq:MI-pole}), respectively.  

\subsubsection{Hourglass}
For the hourglass geometry, we have
\begin{align}
	\mathcal E(\theta) &=h^{\mathcal E\!}(\theta)\log(|\partial A|/\delta) \\
 I(\theta) &= h^{I\!}(\theta)\log(|\partial A|/\delta)
\end{align}
where we omit subleading terms. The prefactors are again protected from the UV cutoff. The LN and MI satisfy the same properties as those found above in Section~\ref{sec:hg-general}, including poles at $\pi$, and monotonicity for the MI \red{and LN, $\partial_\theta h^{I, \mathcal E}(\theta)\geq 0$}. The EE of the hourglass will also contain a logarithmic enhancement $a_\times(\theta) \log(|\partial A|/\delta)$, and the corresponding prefactor is decreasing on $[0,\pi /2]$ and convex, $a_\times''(\theta) \geq 0$, as in Eq.~(\ref{eq:a-times-properties}).

\section{Logarithmic negativity of integer quantum Hall states}\label{sec:method}
We compute the \red{fermionic} LN $\mathcal{E}$ and the MI for various IQH states, including at finite temperature. The single-particle wave function of the $n$-th Landau level (LL) with the eigen-energy $\epsilon_n$ on a torus of area $L_xL_y$ in the Landau gauge $\vec{A}=(0,Bx)$ is: 
\begin{equation}\label{eq:landau}
\phi_{n,k}(x,y)=\frac{1}{\pi^{\frac{1}{4}}\sqrt{2^n n! L_y}}e^{i k y}e^{-\frac{(x+k)^2}{2}}H_n(x+k),
\end{equation}
where $n = 0, 1, 2...$, $k=2\pi m/L_y$ with $m\in \mathbb{Z}$ and $\frac{-L_xL_y}{4\pi}+1\leq m\leq\frac{L_xL_y}{4\pi}$, and we set the units such that the magnetic length $l_B = 1$ and the cyclotron frequency $\hbar\omega_c=1$. We also absorb the zero point energy into the definition of the chemical potential $\mu$, so that LL energies are $\epsilon_n=n$. For each $n$, the degeneracy is $N_d=L_xL_y/2\pi$. IQH states are many-body wave functions constructed from the single-particle wave function (\ref{eq:landau}) with an integer filling fraction $\nu=N_e/N_d$ where $N_e$ is the electron number $N_e$. 

Computing the LN $\mathcal{E}$ for many-body states is generally not an easy task. However, as we will see in the following, it becomes numerically feasible for Gaussian states like IQH states.
 
 \subsection{ Logarithmic negativity for fermionic Gaussian states}\label{sec:LNGaussian}


If the density matrix $\rho_A$ is a Gaussian operator, then so is the density matrix $\rho^{\mathcal T_1}_A$, and thus the normalized composite density operator $\rho_{\times}$ (\ref{eq:composeden}) is also Gaussian.\cite{Shapourian1} For systems with a conserved particle number, the LN $\mathcal{E}$ can be computed from the correlation function $C_{ij}$\cite{Peschel, Shapourian2}: 
\begin{equation} 
C_{ij}=\langle f^{\dagger}_i f_j\rangle=\Tr{(\rho f^{\dagger}_if_j)}=\sum_{n}n_F(\epsilon_n) u^*_n(i)u_n(j) \label{eq:corrFD}
\end{equation}
where $u_n(i)$ is the $n$-th single-particle wave function with eigen-energy $\epsilon_n$, and in the last equality we have restricted ourselves to a thermal state with $n_F(\epsilon_n)=(1+\exp{(\epsilon_n-\mu)/T})^{-1}$ being the Fermi-Dirac distribution function at temperature $T$ and chemical potential $\mu$. The entanglement entropy of subregion $A=A_1A_2$ can then be computed from the eigenvalues $\lbrace\zeta_j \rbrace$ of the correlation matrix $C$ restricted to subregion $A$,\cite{Peschel}
\begin{equation} \label{eq:eig_sumEE}
S(A)=\sum_{j}\left[-\zeta_j\log{\zeta_j}-(1-\zeta_j)\log{(1-\zeta_j)}\right].
\end{equation}
To compute the LN, one needs the composite correlation function $C_{\times}$ associated with the normalized composite density matrix $\rho_{\times}$ (\ref{eq:composeden}). Suppose the covariance matrix of the original density matrix $\rho_A$ (\ref{eq:Dnmaj}) is $\Gamma=\mathbb{1}-2C$, then the covariance matrix for the density matrix defined via the fermionic PT, $\rho^{\mathcal T_1}_A$ and its conjugate,
$(\rho^{\mathcal T_1}_A)^{\dagger}$, can be constructed as
\begin{equation}\label{eq:gammapm}
\Gamma_{\pm}=\begin{bmatrix}
-\Gamma_{11}&\pm i \Gamma_{12}\\
\pm i \Gamma_{21} &\Gamma_{22}
\end{bmatrix}
\end{equation}
where the subindices $1$ and $2$ refer to the subregion $A_1$ and $A_2$, respectively. Following the algebra of the product of Gaussian operators\cite{Fagotti}, one finds that 
\begin{equation}\label{eq:associatedC}
C_{\times} =\frac{1}{2}[\mathbb{1}-\left(\mathbb{1}+\Gamma_+\Gamma_-\right)^{-1}\left(\Gamma_++\Gamma_-\right)].
\end{equation}
\\
As a result, the LN $\mathcal{E}$ can be computed through the spectrum of $C_{\times}$ and $C$:
\begin{equation}
\begin{aligned}
&\mathcal{E}=\sum_{j}\log{\left[\varepsilon^{\frac{1}{2}}_j+\left(1-\varepsilon_j\right)^{\frac{1}{2}}\right]}+\frac{1}{2}\sum_j\log{\left[\zeta^2_j+\left(1-\zeta_j\right)^2\right]}, \label{eq:eig_sum}
\end{aligned}
\end{equation}
where $\varepsilon_j$ are the eigenvalues of the composite correlation matrix $C_{\times}$. Therefore, to compute the LN $\mathcal{E}$ and the MI $I(A_1,A_2)$, one needs the spectrum of the correlation function $ C_{\mathbf{r},\mathbf{r}'}$ and the composite correlation function $C_{\times}$ (\ref{eq:associatedC}).

For IQH states, the electron annihilation operator can be written as $\psi(x,y)=\sum_{n,k}\phi_{n,k}(x,y)c_{n,k}$, where $c_{n,k}$ is the fermionic annihilation operator for the state labelled by $k$ in the $n$th LL. The groundstate wave function at filling $\nu$ is $\vert \Phi_0\rangle=\prod_{k,n<\nu}c^{\dagger}_{n,k}\vert 0\rangle$ with $\vert 0\rangle$ denoting the Fock vacuum. The ground state correlation function is thus
\begin{align}\label{eq:corresum}
C_{\mathbf{r},\mathbf{r}'}=\langle\Phi_0 \vert \psi^{\dagger}(\mathbf{r})\psi(\mathbf{r}')\vert\Phi_0\rangle=\sum_{k,n<\nu}\phi^*_{n,k}(\mathbf{r})\phi_{n,k}(\mathbf{r}')
\end{align}
and the composite correlation function $C_{\times}$ can then be constructed based on Eq.~(\ref{eq:associatedC}).

We develop two independent approaches to compute the spectrum of both the correlation functions $C$ and $C_{\times}$. One is an overlap matrix method in momentum space, and the other is a discretization method in real space.
 
\subsection{Overlap matrix method}\label{sec:overlap}
We first develop an overlap matrix technique to efficiently obtain the fermionic LN. An analogous method has been used to compute the EE of IQH states.\cite{Rodriguez, Sirois} Ref.~\onlinecite{Chang} previously generalized the overlap matrix method for computing the LN defined through the bosonic PT (\ref{eq:LNB}). However, it is very difficult to use such overlap matrix method to study the LN on a general two-dimensional geometry due to its inherent computational complexity: the partially transposed $\rho_{A_1 A_2}^{T_1}$ is non-Gaussian. 
Here, we find a numerically-efficient overlap matrix method to compute the LN defined through the fermionic PT (\ref{eq:FPT}).  

The overlap matrix of subregion $A_1$ is defined as  
\begin{equation} \label{eq:defoverlap}
F^{(1)}_{(n,k), (n', k')}=\int_{A_{1}} d^2\mathbf{r}\ \phi_{n,k}(\mathbf{r})\phi^*_{n',k'}(\mathbf{r}),
\end{equation}
similarly for $A_2$.
The spectrum $\lbrace \zeta\rbrace$ of the correlation matrix $C_{\mathbf{r},\mathbf{r}'}$ on subregion $A_1 A_2$ can be computed from the total overlap matrix, $F^{(12)} = F^{(1)} +F^{(2)}$. On the other hand, obtaining the spectrum $\lbrace \varepsilon\rbrace$ of the composite correlation function $C_\times$, Eq.~(\ref{eq:associatedC}), involves more effort. We use the eigenvalue problem of $\Gamma_{\pm}$ as a starting point to show how the spectrum can be computed through the overlap matrix method:
\begin{equation}\label{eq:overlapGammaP}
\Gamma_{\pm}u=\begin{pmatrix}
-\Gamma_{11}&\pm i\Gamma_{12}\\
\pm i\Gamma_{21}&\Gamma_{22}
\end{pmatrix}\begin{pmatrix}
u^{(1)}\\u^{(2)}
\end{pmatrix}=\lambda_{\pm} \begin{pmatrix}
u^{(1)}\\u^{(2)}
\end{pmatrix},
\end{equation}
where $u^{(1,2)}$ denotes the component of the eigenvector $u$ in subregion $A_{1,2}$. To obtain the spectrum $\lbrace \lambda_{\pm} \rbrace$, we first expand the eigenfunctions $u^{( 1,2)}(\mathbf{r})=\sum_{k,n<\nu}\phi^*_{n,k}(\mathbf{r}) \mathrm{v}^{( 1,2)}_{n,k}$, and using $C_{\mathbf{r},\mathbf{r}'}=\sum_{k,n<\nu}\phi^*_{n,k}(\mathbf{r})\phi_{n,k}(\mathbf{r}') $, the off-diagonal block part becomes
\begin{equation}\label{eq:overlapGamma21}
\begin{aligned}
&\Gamma_{12}u^{( 2)}=-2C_{12}u^{(2)}=-2\int_{A_2} d^2\mathbf{r}' C_{\mathbf{r},\mathbf{r}'}u^{( 2)}(\mathbf{r}')\\
&=-2\sum_{\substack{k, n<\nu\\k', n'<\nu}}\phi^*_{n,k}(\mathbf{r})F^{(2)}_{(n, k),(n', k')}\mathrm{v}^{(2)}_{n',k'}.
\end{aligned}
\end{equation}
By multiplying by $\phi_{m,q}(\mathbf{r})$ and integrating both sides of Eq.~(\ref{eq:overlapGamma21}) \red{on the entire plane}, we obtain
\begin{equation}
\int d^2\mathbf{r}\,  \phi_{m,q}(\mathbf{r}) \Gamma_{12}u^{( 2)}=-2\sum_{k', n'<\nu}F^{(2)}_{(m, q),(n', k')}\mathrm{v}^{( 2)}_{n',k'}.
\end{equation}

Proceeding similarly with the other terms of Eq.~(\ref{eq:overlapGammaP}), the eigenvalue problem $\Gamma_{\pm}u=\lambda_{\pm}u$ in the end can be mapped to the eigenvalue problem of the following overlap matrix:
\begin{equation}
\begin{aligned}
 \begin{pmatrix}
-\left(\mathbb{1}-2F^{\left(1\right)}\right)&\mp 2iF^{\left(2\right)}\\ \mp 2iF^{\left(1\right)} &\left(\mathbb{1}-2F^{\left(2\right)}\right)
\end{pmatrix}\begin{pmatrix}\mathrm{v}^{\left( 1\right)}\\ \mathrm{v}^{\left( 2\right)} \end{pmatrix}=\lambda_{\pm} \begin{pmatrix} \mathrm{v}^{\left( 1\right)}\\ \mathrm{v}^{\left( 2\right)} \end{pmatrix}.\\
\end{aligned}
\end{equation}
This equation is now cast into a finite-matrix eigenvalue problem since the momenta are discrete due to the periodicity in the $y$-direction, 
and we impose a large-momentum cutoff (equivalent to discarding electrons far from the entanglement cut $\partial A$).    

Similarly, we can obtain the spectrum $\lbrace \varepsilon\rbrace$ of the composite correlation function $C_{\times}$ through $C_{\times}\psi_{\times}=\varepsilon \psi_{\times}$, which is equivalent to 
\begin{equation}
\begin{aligned}\label{eq:spectrumCcross}
D^{-1}_{\times}B_{\times}\psi_{\times} &= \left(\Gamma_++\Gamma_-\right)^{-1}\left(\mathbb{1}+\Gamma_+\Gamma_-\right)\psi_{\times} \\ &=\frac{1}{1-2\varepsilon}\psi_{\times}.
\end{aligned}
\end{equation}
where $D_\times = \Gamma_+ + \Gamma_-$, and $B_\times=\mathbb{1} + \Gamma_+ \Gamma_-$. In the momentum basis:
\begin{equation}\label{eq:CcrossA}
D^{-1}_{\times}=\frac{1}{2}\begin{pmatrix}
-\left(\mathbb{1}-2F^{(1)}\right)^{-1} & 0\\0&\left(\mathbb{1}-2F^{(2)}\right)^{-1} 
\end{pmatrix}
\end{equation}
and
\begin{equation}\label{eq:CcrossB}
\begin{aligned}
&B_{\times}=\begin{pmatrix}
\mathbb{1}& 0\\0 & \mathbb{1}
\end{pmatrix}+\\
&\begin{pmatrix}
    -\left(\mathbb{1}-2F^{(1)}\right)&-2iF^{(2)}\\-2iF^{(1)} &\mathbb{1}-2F^{(2)}
\end{pmatrix}\begin{pmatrix}
-\left(\mathbb{1}-2F^{(1)}\right)&2iF^{(2)}\\2iF^{(1)} &\mathbb{1}-2F^{(2)}
\end{pmatrix}.
\end{aligned}
\end{equation}
The  spectrum $\lbrace \varepsilon\rbrace$  can thus be obtained through Eqs.~(\ref{eq:spectrumCcross})-(\ref{eq:CcrossB}).

\subsection{Real space discretization method}
The correlation function $C_{\mathbf{r},\mathbf{r}'}$ is defined on continuous real space. To obtain its spectrum $\lbrace \zeta\rbrace$ on a subregion $A$ in real space, we need to solve a functional eigenvalue problem:
\begin{equation}\label{eq:eigprob}
\int_A d^2\mathbf{r}' C_{\mathbf{r},\mathbf{r}'}u(\mathbf{r}')=\zeta u(\mathbf{r}).
\end{equation}
Here we take the thermodynamic limit $L_x,L_y\rightarrow \infty$, and the summation in Eq.~(\ref{eq:corresum}) can then be replaced by an integral. At $T=0$, we focus on fillings $\nu=1$ and $\nu=2$. From Eq.~(\ref{eq:corresum}), the correlation functions are the following:
\begin{widetext} 
\begin{equation}\label{eq:correinf}
\begin{aligned}
C_{\mathbf{r},\mathbf{r}'}=\frac{1}{4\pi}e^{-\frac{1}{4}|\mathbf{r}-\mathbf{r}'|^2-\frac{i}{2}(x+x')(y-y')} \times\begin{cases}
2 , & \nu=1\\
4-|\mathbf{r}-\mathbf{r}'|^2 , & \nu=2.
\end{cases}
\end{aligned}
\end{equation}
\end{widetext}

We solve the functional eigenvalue problem (\ref{eq:eigprob}) by discretizing the continuous real space, that is, we solve it on a finite partition  $\lbrace\mathbf{r}_i \rbrace( \, i=1,2,\ldots ,N)$ of the subregion $A$. After discretization, the integral is replaced by a Riemann sum, and the functional eigenvalue problem (\ref{eq:eigprob}) becomes 
\begin{equation}\label{eq:eigendiscrete}
\begin{aligned}
\int_A d^2\mathbf{r}' C_{\mathbf{r},\mathbf{r}'} u(\mathbf{r}')&\cong\sum_{j\in A}\Delta x_j
\Delta y_j \widetilde{C}_{ij}\widetilde{u}(j)\\
&=\sum_{j\in A}a^2\widetilde{C}_{ij}\widetilde{u}(j)=\zeta\widetilde{u}(i),
\end{aligned}
\end{equation}
where we choose a square lattice with spacing $\Delta x_j=\Delta y_j=a$, and denote the objects on the discrete lattice  $\lbrace\mathbf{r}_i \rbrace$ by the over-tilde symbol as $\widetilde{C}_{ij}\equiv C_{\mathbf{r}_i\mathbf{r}_j}$ and $\widetilde{u}({i})\equiv u(\mathbf{r}_i)$. From Eq.~(\ref{eq:eigendiscrete}), we see that to solve the spectrum $\lbrace \zeta\rbrace$ of the correlation function $C_{\mathbf{r},\mathbf{r}'}$, one needs to solve the eigenvalue problem of the matrix $a^2\widetilde{C}_{ij}$ instead of  $\widetilde{C}_{ij}$. 
Moreover, since the discrete version of the Dirac delta function is
$\delta(\mathbf{r}'-\mathbf{r}'')\rightarrow \mathbb{1}_{ij}/a^2$, where $\mathbb{1}$ is the identity matrix,  the discrete version of the inverse function of $A^{-1}_{\mathbf{r}',\mathbf{r}''}$ should include an extra prefactor $1/a^4$,
$A^{-1}_{\mathbf{r}',\mathbf{r}''}\rightarrow \widetilde{A}^{-1}_{ij}/a^4$,
so that the relation 
\begin{equation}
\int_A d^2\mathbf{r}' A_{\mathbf{r},\mathbf{r}'}A^{-1}_{\mathbf{r}',\mathbf{r}''}=\delta(\mathbf{r}-\mathbf{r}'')
\end{equation}
holds in its discrete form. 
Following these rules, the discrete version of the composite correlation function $C_{\times}$ is
\begin{equation}\label{eq:associatedCAP}
\widetilde{C}_{\times,ij} =\frac{1}{2a^2}\left[\mathbb{1}_{ij}-\sum_{k}\left(\frac{ \mathbb{1}}{a^2}+a^2\widetilde{\Gamma}_+\widetilde{\Gamma}_-\right)_{ik}^{-1}(\widetilde{\Gamma}_{+}+\widetilde{\Gamma}_{-})_{kj}\right]\! ,
\end{equation}
and its spectrum can be computed as in Eq.~(\ref{eq:eigendiscrete}).

The eigenvalues of $\tilde C_{ij}$ and $\tilde C_{\times,ij}$ depend on the lattice spacing $a$. To avoid such cut-off dependence, we extrapolate our results for the LN and MI to the $a\to 0$ limit. Fig.~\ref{fig:negparahourfiniteL10} in Appendix (\ref{appendix:data}) shows an example of the finite-size scaling at $\theta=0.3\pi$. 
\section{Integer quantum Hall groundstates}\label{sec:tripartieZeroT}

In this section, we present our numerical results based on the overlap matrix method for the LN and MI of IQH groundstates on various tripartite geometries with corners. In Appendix~\ref{appendix:data}, we compare these with results from the real space discretization method. Both methods
agree, but the overlap matrix method gives superior precision, at a reduced computational cost.

\subsection{Adjacent geometry}\label{sec:adj}
\begin{figure}
\centering
\includegraphics[scale=0.23]{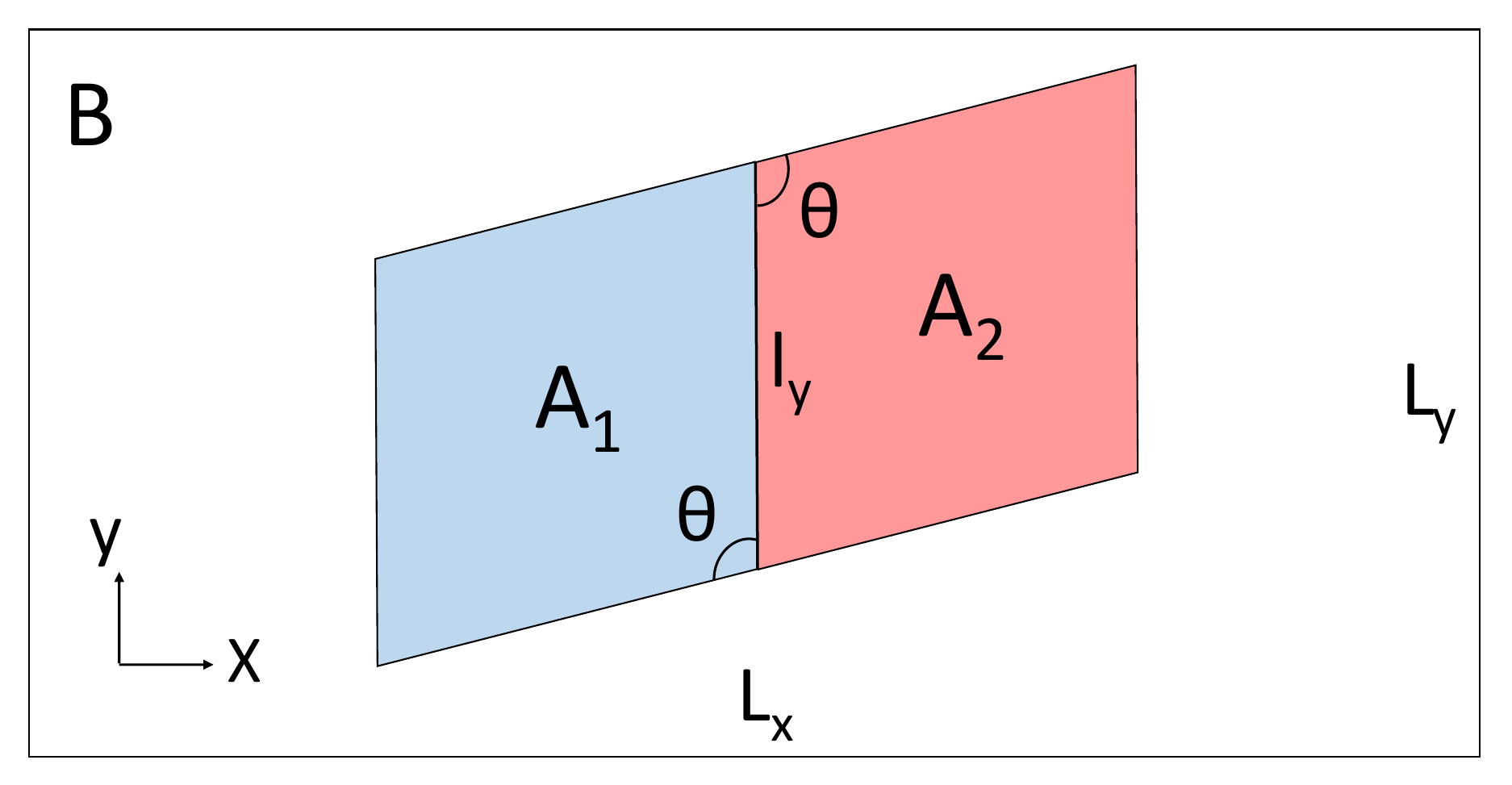}
\caption{Tripartite adjacent geometry. Subregions $A_1$ and $A_2$ share a boundary of length $l_y$, and have two pairs of touching corners.}
\label{fig:geometry3}
\end{figure}

First, we compute the LN and the MI for a geometry with adjacent parallelograms as shown in Fig.~\ref{fig:geometry3} for various angles $\theta$. We find that both the LN and the MI obey the following form in the thermodynamic limit: 
\begin{align} 
&\mathcal{E}(A_1,A_2)=c_{\mathcal{E}}l_y -2b(\theta), \label{scaling}\\
&I(A_1,A_2) =2c_{1}l_y-2b^{I\!}(\theta)\label{MIscaling}
\end{align}
where $l_y$ is the length of the boundary shared by $A_1$ and $A_2$, and $b(\theta)\equiv b(\theta,\pi-\theta)$,
where we use a simplified notation compared to the
the general adjacent corner term with angles $\theta_1,\theta_2$. 
For the MI, we have a simple expression in terms of the single-corner function, $b^{I\!}(\theta)=a(\theta)+a(\pi-\theta)$. 
The factors of 2 in the subleading terms come from the 2 adjacent pairs. 

Tables~(\ref{Tab:slope}) and (\ref{Tab:adjancent}) list the values of the boundary law coefficients and the subleading corner functions. The boundary law coefficient $c_{\mathcal{E}}$ of the LN is just the same as the boundary law coefficient of $S_{1/2}(A_1)$ since the boundary law should be insensitive to the geometry, and on bipartite geometry the LN  $\mathcal{E}$ is just the same as $S_{1/2}(A_1)$. 

\begin{figure}
\centering
\includegraphics[scale=0.60]{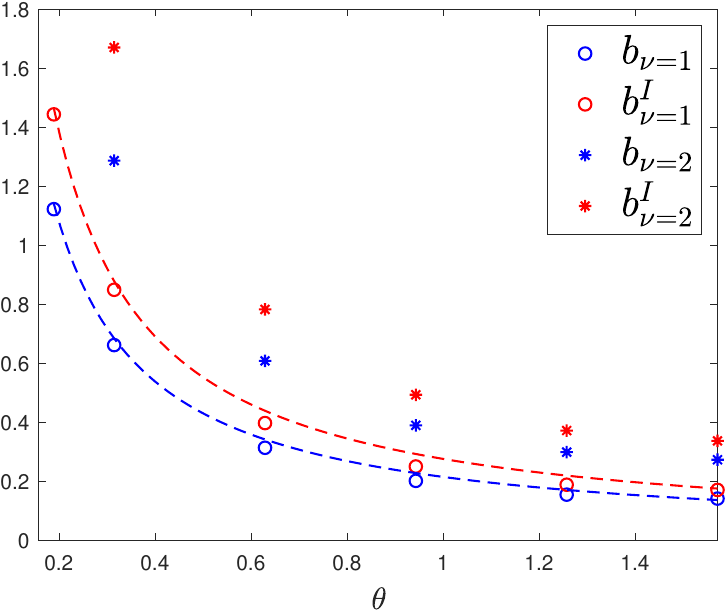}
\caption{\label{fig:adjresults} Angle dependence of the subleading LN term, $b$, and subleading MI term, $b^I$, at fillings $\nu = 1,2$ on the adjacent geometry. The curves show the small angle behaviour: $k_{\rm adj}/\theta$ for the LN, and $\kappa_{\rm adj}/\theta$ for the MI.
The small-angle coefficients are given in Table~\ref{Tab:kappa}.
}
\end{figure}

\begin{figure}
    \centering
    \includegraphics[scale=0.42]{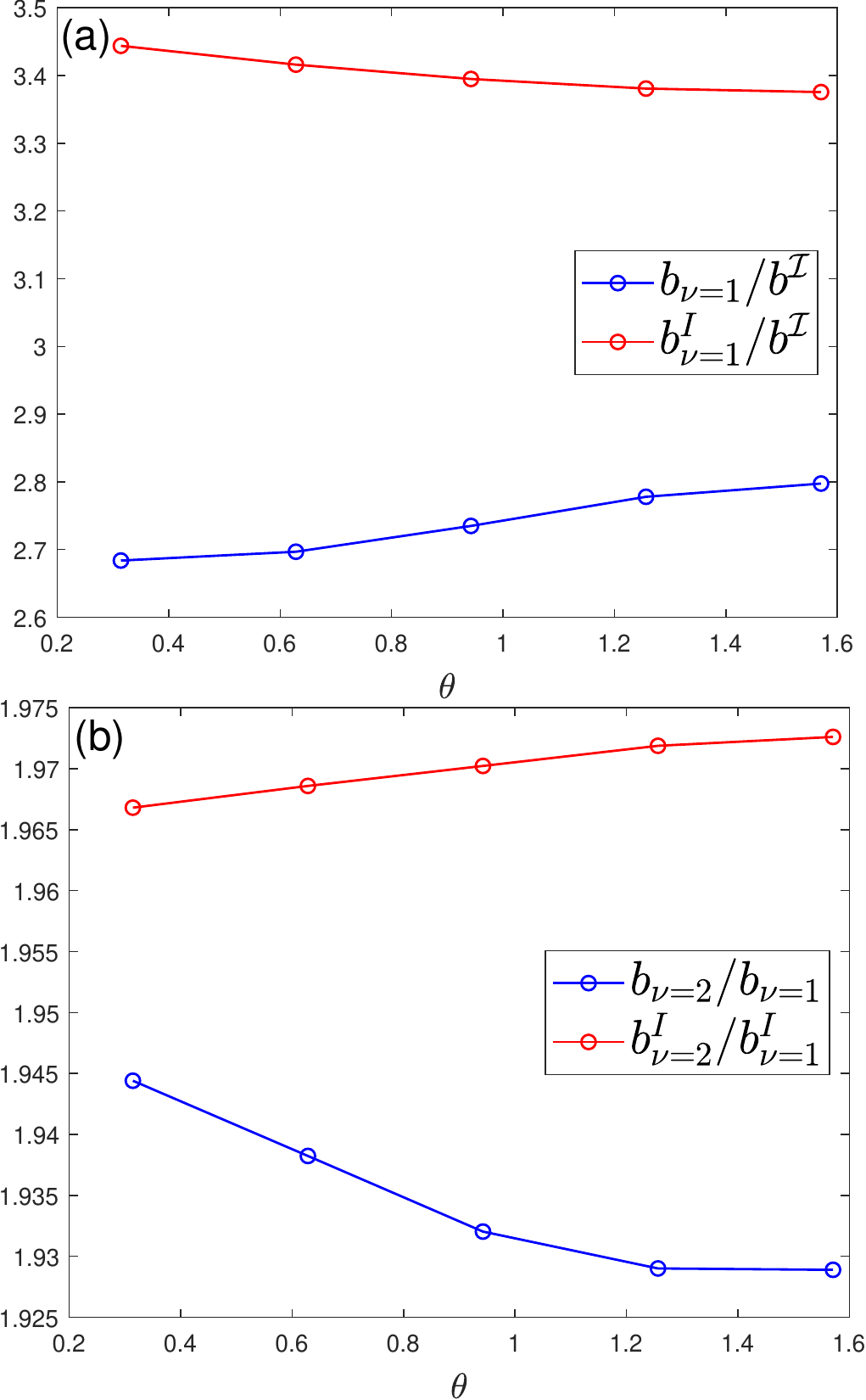}
    \caption{\label{fig:negcorner5} Ratios of corner terms for the adjacent geometry. (a) The MI corner term  $b^I(\theta )$, and the LN corner term $b(\theta)$ divided by the fluctuation corner term $b^\mathcal{I}(\theta)$. (b) The ratios of the LN corner term $b(\theta )$ and MI corner term $b^I(\theta )$ between filling $\nu = 1$ and $\nu = 2$.} 
\end{figure}

The LN and MI corner functions are shown in  Fig.~\ref{fig:adjresults} for fillings $\nu=1,2$. For charge fluctuations, using the analogue of Eq.~(\ref{eq:a-fluc}) (see also [\onlinecite{Estienne2}]), we have
\begin{align}
\label{fitting}
b^{\mathcal I\!}(\theta)=\frac{\nu}{4\pi^2}(2 + (\pi-2\theta)\cot{\theta}).
\end{align}
In all cases, we observe a $1/\theta$ divergence at small angles, in agreement with the general results given above, Eqs.~(\ref{eq:LN-pole})-(\ref{eq:MI-pole}). For the LN at filling $\nu=1$, 
we numerically determine that the coefficient of $1/\theta$ is $k_{\rm adj}= 0.215$, as presented in Table~\ref{Tab:kappa} along with the other small angle coefficients. As the angle increases from zero, we find that the LN, MI and fluctuations decreases in a monotonous fashion, reaching their minimum at $\pi/2$, in agreement with the general findings in Section \ref{sec:adj-general}. In particular, the behaviour about the minimum satisfies Eq.~(\ref{eq:adj-smooth}).
Fig.~\ref{fig:negcorner5}(a) shows that the ratios of $b(\theta)$ and $b^I(\theta)$ to the charge fluctuation corner function $b^\mathcal{I}(\theta )$. We note that the ratio shows little dependence on the angle, indicating that all three quantities share almost the same geometrical dependence. 
In Fig.~\ref{fig:negcorner5}(b), we show the ratio of filling $\nu=2$ to filling $\nu=1$ for the LN and MI. The ratios again vary slowly with the angle, and hover near 2. The naive expectation that having two filled Landau levels should give twice the contribution of the $\nu=1$ groundstate is almost born out, but only holds exactly for mutual fluctuations. The LN shows the strongest deviation from 2. It would be of interest to understand why this is so.

\subsection{Hourglass geometry}\label{sec:hg}
\vspace{-2mm}
\begin{figure}
\centering
\includegraphics[scale=0.24]{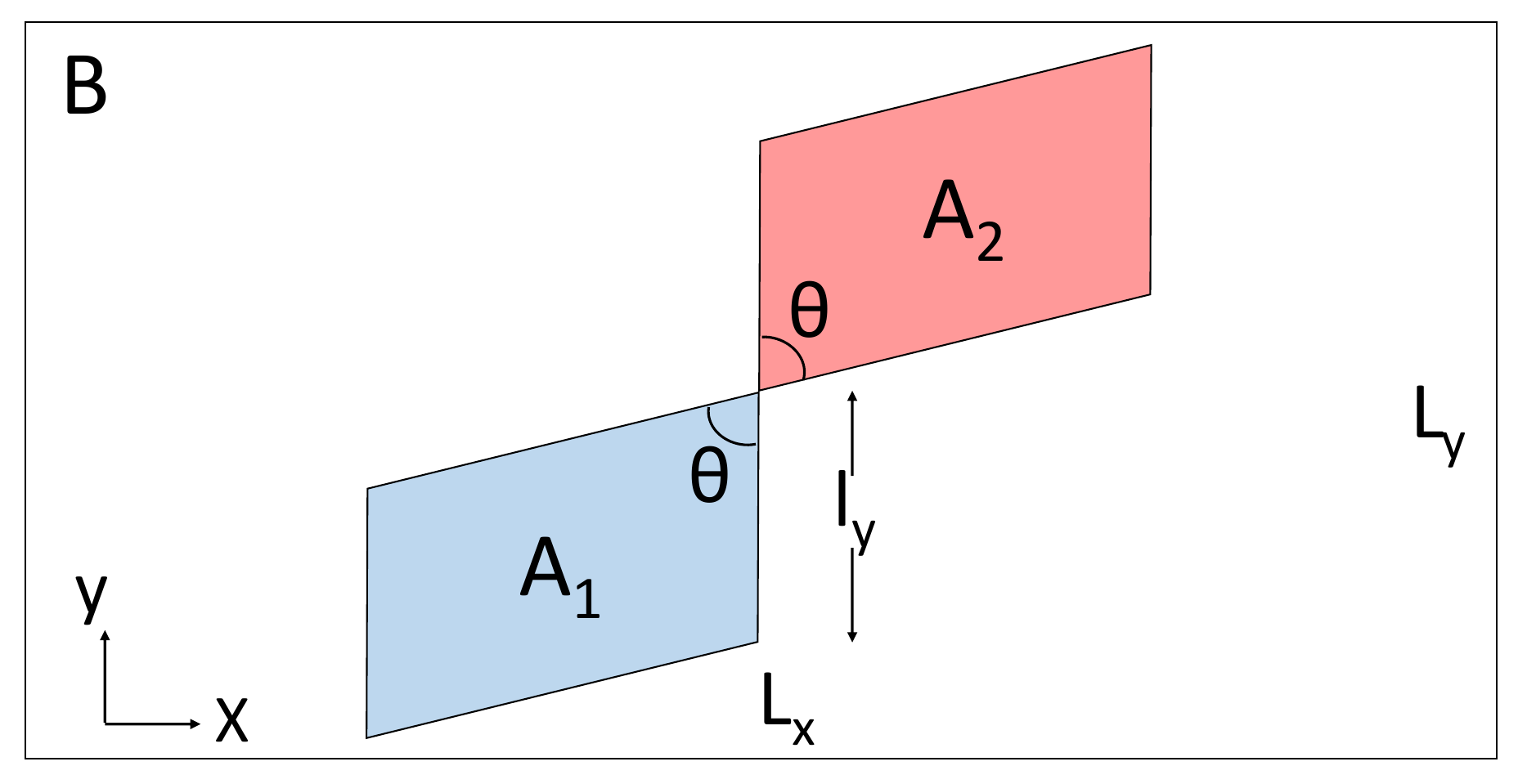}
\caption{Tripartite hourglass geometry. Subregions $A_1$ and $A_2$ touch at a single point, with an hourglass corner of angle $\theta$.}
\label{fig:geometryhourglass}
\end{figure}

\begin{figure}
\centering
\includegraphics[scale=0.69 ]{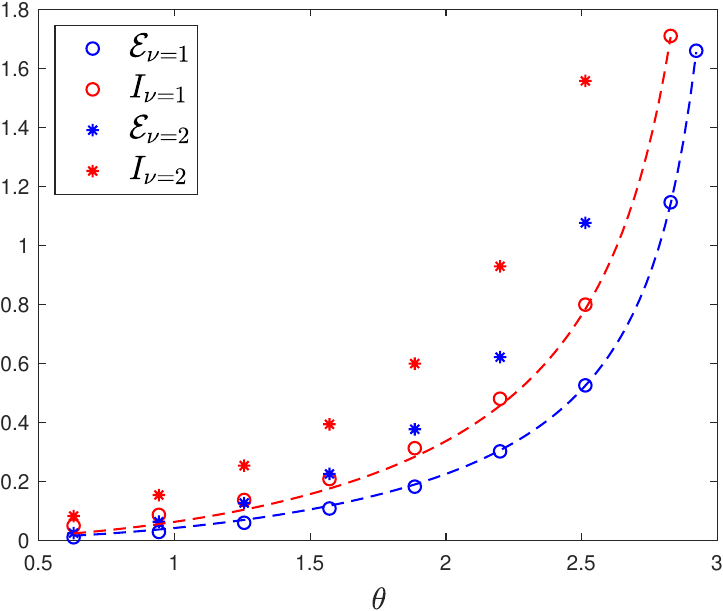}
\caption{\label{fig:negmutparahournu1L10}  The LN $\mathcal{E}$ and the MI $I$ at various angles for fillings $\nu=1,2$ in the hourglass geometry. The two dashed lines correspond to the hourglass function for mutual charge fluctuations $\mathcal{I}(\theta)$ (\ref{eq:fluctuationhour}), but normalized such that it reproduces the exact small-angle behavior for the LN and MI at $\nu=1$, $\frac{\{k_\times, \kappa_\times\} }{\pi}(1-\theta\cot{\theta})$,
respectively;
the corresponding small-angle coefficients appear in Table~\ref{Tab:kappa}. }
\end{figure}

We now turn to tip-touching corners. The calculations are done on the parallelogram hourglass geometry shown in Fig.~\ref{fig:geometryhourglass}. The hourglass geometry is of particular interest because the subregions $A_1$ and $A_2$ only touch at a point, characterized by an angle $\theta$. Thus, there is no boundary law between the two subregions, and we can focus on the geometric corner contribution to the LN. The MI was previously studied at $\nu=1$ for $\theta=\pi/2$.\cite{Sirois} As for the adjacent geometry, we also compare the LN with the corner function $\mathcal{I}(\theta)$ of charge fluctuations on an hourglass geometry:\cite{Berthiere}
\begin{equation}\label{eq:fluctuationhour}
\mathcal{I}(\theta)=\frac{\nu}{4\pi^2}(1-\theta\cot{\theta}).
\end{equation}
The angle dependence of the LN $\mathcal{E}$ and the MI for $\nu=1,2$ are shown in Fig.~\ref{fig:negmutparahournu1L10}.
The LN and MI vanish at small angles, in agreement with the general analysis of Section \ref{sec:hg-general}. We note that the LN decays faster than the MI. As the angle increases towards $\pi$, a pole emerges for the LN and MI, as given in Eqs.~(\ref{eq:LN-pi})-(\ref{eq:MI-pi}). For the LN, we find that the coefficient (residue) is $k_\times=0.369(1)$. For the MI, we can use the relation to the single-corner coefficient $\kappa$, Eq.~(\ref{eq:kappax}), to get $\kappa_\times=0.552$.\cite{Sirois} The coefficients of the divergences (pole residues) for the LN and MI are summarized in Table~\ref{Tab:kappa}. The dashed lines correspond to the mutual fluctuations function $(1-\theta\cot\theta)/\pi$ with prefactor $k_\times$ for the LN, and $\kappa_\times$ for the MI. These thus accurately capture the divergence at $\pi$. We see that they also provide a reasonable estimate at smaller angles, without any additional fitting parameters. However, deviations can be seen indicating that the LN and MI have a distinct angle dependence compared with the mutual fluctuations, \red{but it is noteworthy that the agreement is surprisingly good in the case of the LN for all angles under consideration. It would be interesting to understand why these two very different quantities possess such a similar angle dependence beyond the general constraints established in Section~\ref{sec:general}.}    

\begin{table}[!htbp]
\caption{The pole residues of the corner functions for the  LN and MI on the single-corner, adjacent corners and hourglass geometries at filling $\nu=1$. The value of $\kappa_{\rm adj}$, the residue of the corner function for the MI on the adjacent geometry, is based on Ref.~\onlinecite{Sirois}. }\label{Tab:kappa}
\begin{center}
\scalebox{1}{
\begin{tabular}{ |ccc ccc ccc  ||ccc ccc ccc|  }
 \hline
 \multicolumn{9}{|c||}{LN}&\multicolumn{9}{c|} {MI}\\
 \hline
& $\kappa^{\mathcal{E}}\!=\kappa_{1/2}$ &&& $k_{\rm adj}$  & &&$k_{\times}$ & && $\kappa^{I}=2\kappa$  & & &$\kappa_{\rm adj}=\kappa $&& & $\kappa_{\times}=2 \kappa$&\\
& 0.475 &&&0.215 & & &0.369& & &0.552 &&&0.276 & & &0.552&\\
 \hline
\end{tabular}
}
\end{center}
\end{table} 

\section{Integer quantum Hall states at finite temperature}\label{sec:LNfiniteT}

The LN is a good measure of entanglement for quantum mixed states since it captures only quantum correlations as opposed to the MI. As such, the LN is well-suited to study entanglement at finite temperature. In this section, we study the finite temperature LN for both the hourglass and adjacent geometries with corners of angle $\pi/2$, and compare our findings with the MI and mutual fluctuations.  

At finite temperature $T>0$, the overlap matrix method requires more LLs, increasing the matrix size, and gradually becomes numerically untractable. Therefore, unless the temperature is very small, we shall use the real space discretization method to compute the finite temperature LN. To do this, we include the Fermi-Dirac distribution as in Eq. (\ref{eq:corrFD}) so that the real space correlation function is  $C_{\mathbf{r},\mathbf{r}'}=\sum_{n,k}n_F(\epsilon_n)\phi^*_{n,k}(\mathbf{r})\phi_{n,k}(\mathbf{r}')$. We work in the grand canonical ensemble, where we solve the chemical potential $\mu(T)$ self-consistently at a given temperature $T$  by fixing the average filling, $\nu$. Summing over all the contributing LLs requires considerable numerical efforts at high temperatures. As such, we only report the finite temperature results up to temperatures on the order of $T = 2$ (in units of the cyclotron energy). We separately explore the high temperature behaviour of the LN by working in the limit $T \rightarrow{\infty}$ in Section~\ref{sec:highT}.

\begin{figure}
    \centering
    \includegraphics[scale=0.42]{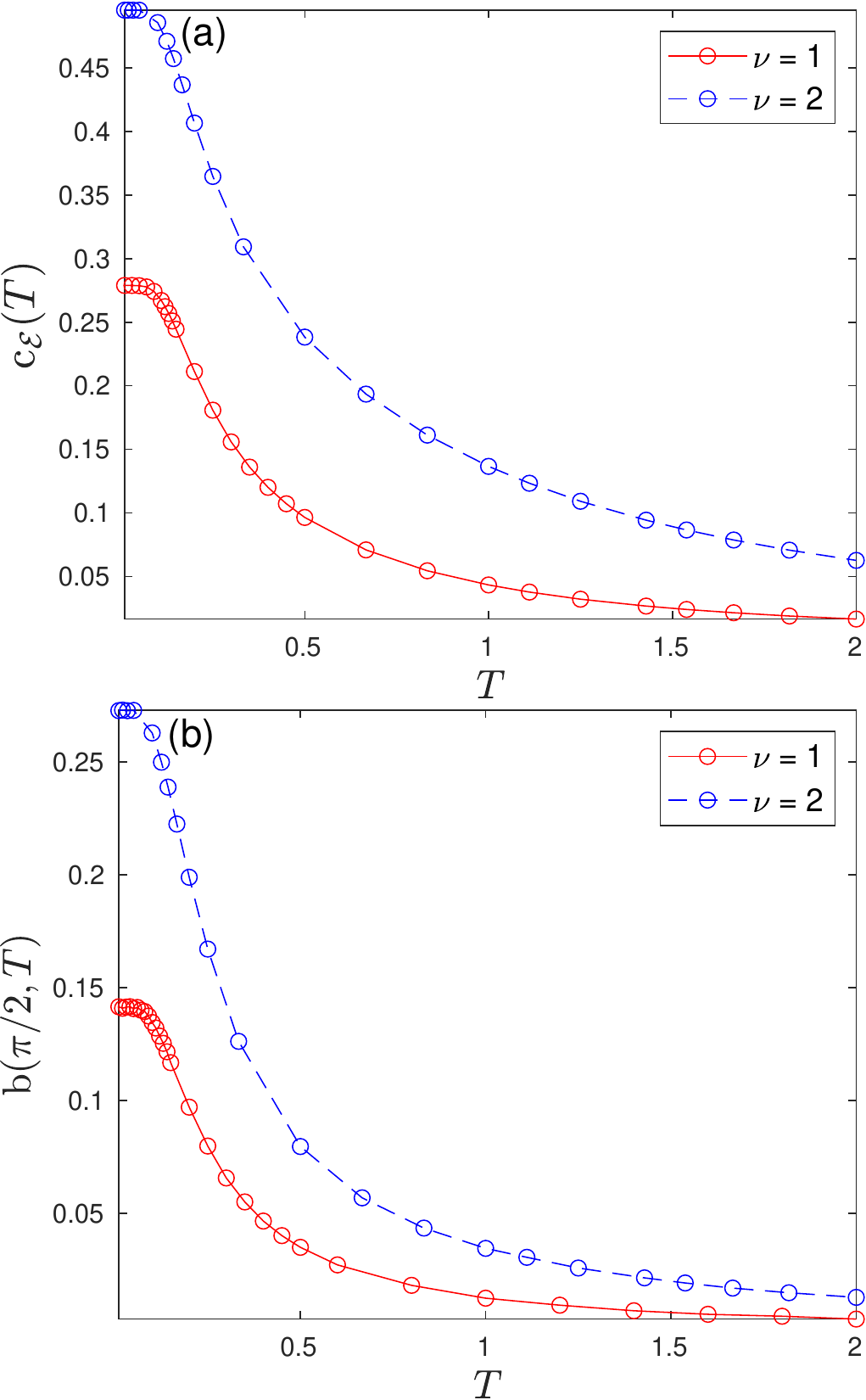}
    \caption{\label{fig:adjacent_temp} Finite temperature LN at average fillings $\nu = 1,2$ on the adjacent geometry. The shared boundary $l_y$ is varied to perform a linear regression and obtain the boundary law coefficient, $c_\mathcal{E}(T)$ and corner term, $b(\pi/2, T)$. We used $L_x= L_y = 40$. (a) Boundary law coefficient as a function of temperature, (b) corner term $b(\pi/2, T)$. }
\end{figure}

\begin{figure}
    \centering
     \includegraphics[scale=0.6]{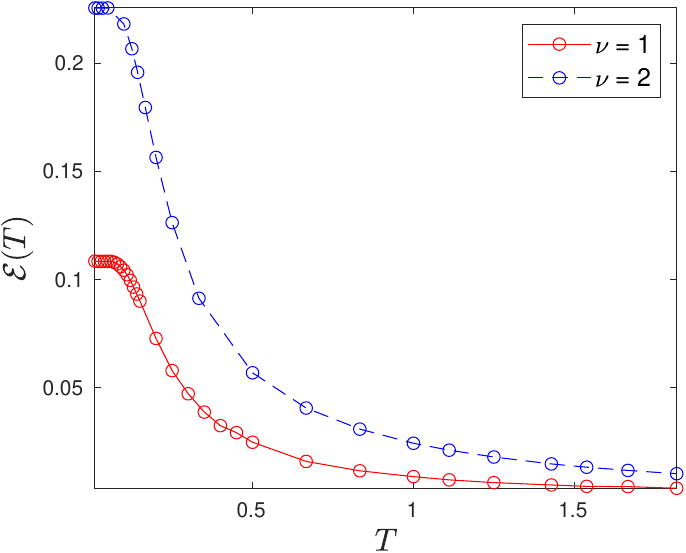}
    \caption{Finite temperature LN  $\mathcal{E}(T)$ at average fillings $\nu =1,2$ on the hourglass geometry at touching angle $\pi/2$ with $l_y = 10$. We used $L_x= L_y = 40$.  }
    \label{fig:hourglass_temp}
\end{figure}

For the adjacent geometry, we find that at finite temperatures the LN still obeys a boundary law with a subleading corner contribution:
\begin{equation}\label{eq:tempAreaLaw}
\mathcal{E}(T) = c_\mathcal{E}(T) l_y - 2b(\pi/2 , T)
\end{equation}
\red{where we have explicitly indicated the temperature dependence.} 
The $T$ dependence of the boundary law coefficient, and of the subleading term at average fillings $\nu = 1,2$ are shown in Fig.~\ref{fig:adjacent_temp}. We also plot the LN for the hourglass geometry in Fig.~\ref{fig:hourglass_temp}. The finite temperature LN for the adjacent and hourglass geometries share similar features. 
Namely, they both plateau at low temperatures until they start decreasing at a small temperature on the order $T \approx 0.1$ and then decay towards zero, which indicates the loss of entanglement as the system heats up. The drop is quite abrupt. For instance, when the temperature reaches the cyclotron gap $T=1$, the LN in hourglass geometry drops to 4\% of its $T=0$ value.
The low-$T$ regime is studied in more detail in the next subsection. 
Similar features also appeared when studying the thermal charge fluctuations of IQH states~\cite{Estienne} or the bosonic LN of harmonic oscillators chains,\cite{Werner} for example. 

\subsection{Low temperatures} \label{sec:lowT}
\begin{figure}
    \centering
   
    \includegraphics[scale=0.4]{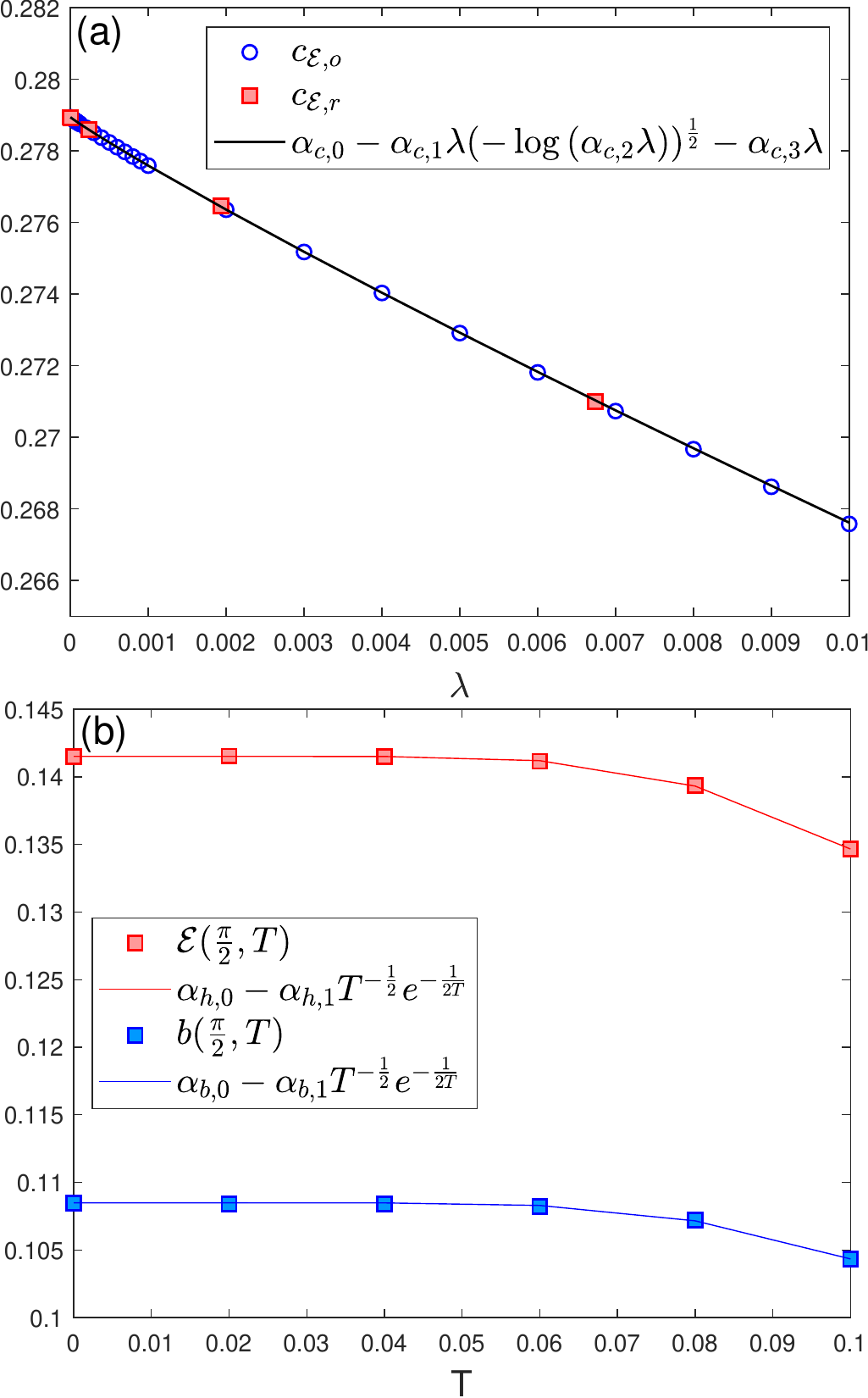}
    
    
    \caption{(a) The boundary law coefficients of the LN computed through the overlap matrix ($c_{\mathcal{E},o}$ denoted as blue dots) and the real space discretization ($c_{\mathcal{E},r}$ denoted as red squares) methods, and the fitting function (black curve) with parameters $(\alpha_{c,0},\alpha_{c,1},\alpha_{c,2},\alpha_{c,3})=(0.2789,  0.4481, 1.2192,0.1916)$  versus $\lambda$. (b) The subleading term of LN on the adjacent geometry at angle $\pi/2$ (blue) and the LN on the hourglass geometry at  touching angle $\pi/2$ (red) by setting $\lambda=e^{-\frac{\beta}{2}}$. The numerical data from the real space discretization method and the fitting functions with parameters $(\alpha_{b,0},\alpha_{b,1})=(0.1415, -0.4547)$ and  $(\alpha_{h,0},\alpha_{h,1})=(0.1085, -0.2753)$  versus $\lambda=e^{-\frac{\beta}{2}}$  are denoted by squares and curves, respectively. }
    \label{fig:LN_lowT}
\end{figure}

In this section, we analyze the LN in the low temperature limit $\beta=1/T\gg 1$. 
The chemical potential is nearly constant  $\mu\cong 1/2 $, so that one can expand the Fermi-Dirac distribution in 
terms of the small parameter
\begin{align}
    \lambda=e^{-\beta/2}\ll  1
\end{align}
The two leading terms in the correlation function \eqref{eq:corrFD} read
\begin{align}
C_{\mathbf{r},\mathbf{r}'} =\Tr{(\rho f^{\dagger}_{\mathbf{r}}f_{\mathbf{r}'})} = (1 - \lambda)     
C_{0,\mathbf{r},\mathbf{r}'}+\lambda C_{1,\mathbf{r},\mathbf{r}'}
\end{align}
where $C_n$ denotes the correlation function in the $n$-th LL. Based on the correlation function above, the LN can be computed through the overlap matrices defined in Appendix~(\ref{appendix:lowToverlap}). For IQH states, our numerical data in the region $\lambda\ll 1$ show that the LN in both the adjacent and hourglass geometry receives exponentially small corrections, that is 
\begin{equation}
\mathcal{E}_{\text{low }T} \cong \mathcal{E}_{T = 0} -\lambda\,\Delta\mathcal{E}(\lambda)+\cdots
\label{eq:LNlowTsupp}
\end{equation}
We observe that our data is consistent with the following slowly varying term: $\Delta \mathcal E(\lambda)\sim (-\log{\lambda})^{\frac{1}{2}}=(T/2)^{1/2}$ in the low temperature limit $\lambda\rightarrow 0$. 
Fig.~\ref{fig:LN_lowT}(a) shows the boundary law coefficient of the LN, $c_{\mathcal{E},o}$ and $c_{\mathcal{E},r}$, computed through the overlap matrix and real space discretization method, respectively, versus $\lambda$. We find that the data is well-described by the fitting function $\alpha_{c,0}-\alpha_{c,1}\lambda(-\log{(\alpha_{c,2}\lambda}))^{\frac{1}{2}}-\alpha_{c,3}\lambda$ with $(\alpha_{c,0},\alpha_{c,1},\alpha_{c,2},\alpha_{c,3})=(0.2789, 0.4481, 1.2192, 0.1916)$ which approaches $\alpha_{c,0}-\alpha_{c,1}\lambda(-\log{(\lambda}))^{\frac{1}{2}}$ in the low temperature limit $\lambda\rightarrow 0$, with an exponentially small correction due to temperature. This robustness at small temperatures is natural given the (cyclotron) gap. Exponentially small thermal corrections were also observed for the corner term of charge fluctuations.\cite{Estienne}
Based on this observation, we also try to fit the subleading term of the LN on the adjacent geometry, $b(\pi/2,T)$, 
as well as the LN on the hourglass geometry computed through the real space discretization method with the fitting function $\alpha_{0}-\alpha_{1}\lambda(-\log{\lambda})^{\frac{1}{2}}$ as shown in Fig.~\ref{fig:LN_lowT}(b). We observe similar exponential suppression at low temperature.

\begin{figure}
    \centering
    \includegraphics[scale=0.55]{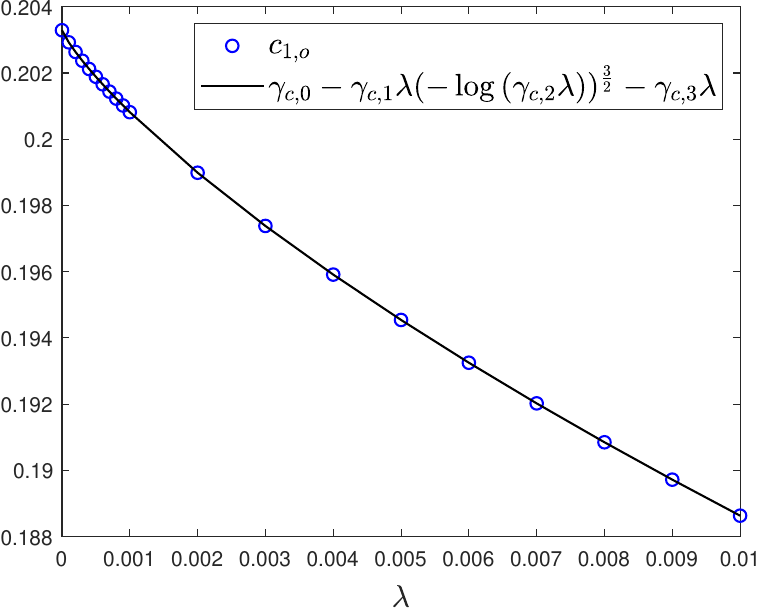}
    \caption{The boundary law coefficient of the entanglement entropy $c_{1}$ (blue dots) computed with the overlap matrix method  versus $\lambda=e^{-\frac{\beta}{2}}$, and the fitting function (black curve) with parameters $(\gamma_{c,0},\gamma_{c,1},\gamma_{c,2},\gamma_{c,3})=(0.2033,    0.1167,  0.5928, 0.1109)$. The boundary law coefficient of the MI is twice as large, $2c_1$. }
    \label{fig:MI_lowT}
\end{figure}

From these results, we see that our data obeys (\ref{eq:LNlowTsupp}), i.e.\ the LN only receives exponentially small negative thermal corrections due to the gap in the spectrum of the IQH system. We leave the rigorous proof of Eq.~(\ref{eq:LNlowTsupp}) for future work. We also point out that such an exponentially negative correction is a common feature in IQH states, and is also present for the boundary law coefficient of the MI as shown in Fig.~\ref{fig:MI_lowT}, and for the charge fluctuation corner term.\cite{Estienne} For the boundary law coefficient of the MI, the thermal correction is $-2\gamma_{c,1}\lambda (-\log\lambda)^{3/2}<0$, which is parametrically stronger than what we obtained for the LN, which had a power 1/2 instead of 3/2. Thus, the MI decreases faster with temperature at asymptotically low $T$ compared with the LN.

\subsection{High temperatures} \label{sec:highT}

In the high temperature limit, thermal fluctuations wash out quantum entanglement. Although we cannot achieve very high temperatures numerically by using the correlation function (\ref{eq:corresum}), we can study the limit $T \rightarrow{\infty}$
analytically. Indeed, the chemical potential is large and negative such that the Fermi-Dirac distribution reduces to the Boltzmann distribution, and the infinite sum of the energy levels $\epsilon_n$ can be evaluated exactly using the integral representation of Hermite polynomials as we show in Appendix~\ref{appendix:highT}. The correlation function is then given by Eq.~(\ref{eq:corr_sum}) in  Appendix~\ref{appendix:highT}, where the dependence on the filling $\nu$ is captured by the chemical potential. 

We can simplify the correlation function further by working in the thermodynamic limit where the momentum summation becomes an integral. As shown in Appendix~\ref{appendix:highT}, the correlation function at $\nu = 1$ has the following simple form when $\beta \ll 1$,
\begin{equation}\label{eq:highTcorr_limit}
C_{\mathbf{r},\mathbf{r}'} \approx \frac{1}{2\pi}e^{-\frac{1}{2\beta}|\mathbf{r}-\mathbf{r}'|^2-\frac{i}{2}(x+x')(y-y')}.
\end{equation} 
In the limit $\beta \rightarrow 0$ , $C_{\mathbf{r},\mathbf{r}'}$ vanishes everywhere except when $\mathbf{r}=\mathbf{r}'$, that is, $C_{\mathbf{r},\mathbf{r}'}$ becomes ultra-local. We prove in Appendix~\ref{appendix:LNnocorrelation} that under this condition the LN vanishes.

As mentioned, at finite temperatures, the overlap matrix method becomes numerically unfeasible because of the many LLs involved. However, at high temperatures $\beta\ll 1$, the correlation function can be approximated as Eq.~(\ref{eq:highTcorr_limit}), and based on that we can develop an overlap matrix method adapted for high temperatures to  numerically compute the thermal entropy and the LN, as described in Appendix~(\ref{appendix:OverlapHighT}). At $\beta\rightarrow 0$, the spectrum of Eq.~(\ref{eq:highTcorr_limit}) on a torus can even be analytically solved, and the leading term of the thermal entropy of subregion $A$ can also be computed through Eq.~(\ref{eq:eig_sumEE}) (See Appendix~\ref{appendix:OverlapHighT} for details):
\begin{equation}\label{eq:thermalEE}
S_{A}(T) \cong |A| \frac{\log T}{2\pi}, 
\end{equation} 
which shows that the EE obeys a volume law. 

\begin{figure}
    \centering
    \includegraphics[scale=0.55]{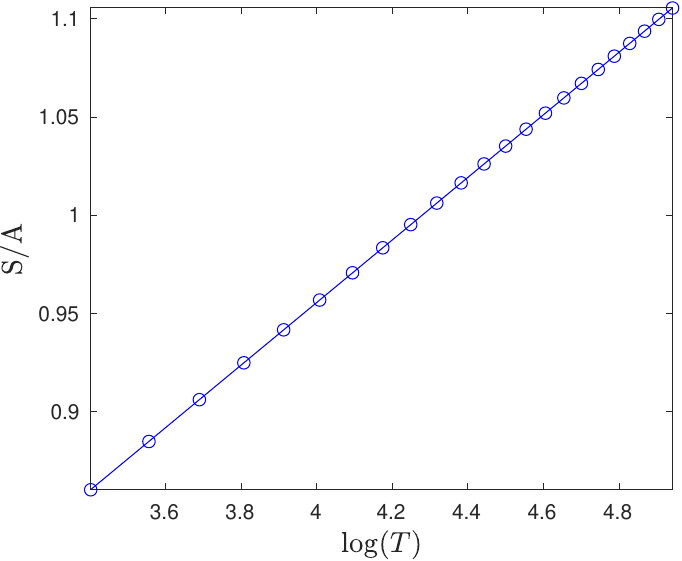}
    \caption{The thermal entropy at high temperature scales logarithmically with temperature. The data points are computed at average filling $\nu = 1$ using the high temperature overlap matrix technique on an adjacent geometry where the length of the boundary of subregion $A$ is the same as the boundary of the whole system ($l_y = L_x = L_y = 285$ and $l_x = 10$). The slope of the fitting line $c=0.1590(1)$ is indeed close to the analytical result $1/2\pi$.  }
    \label{fig:EE_hightemp}
\end{figure}

\begin{figure}
    \centering
    \includegraphics[scale=0.57]{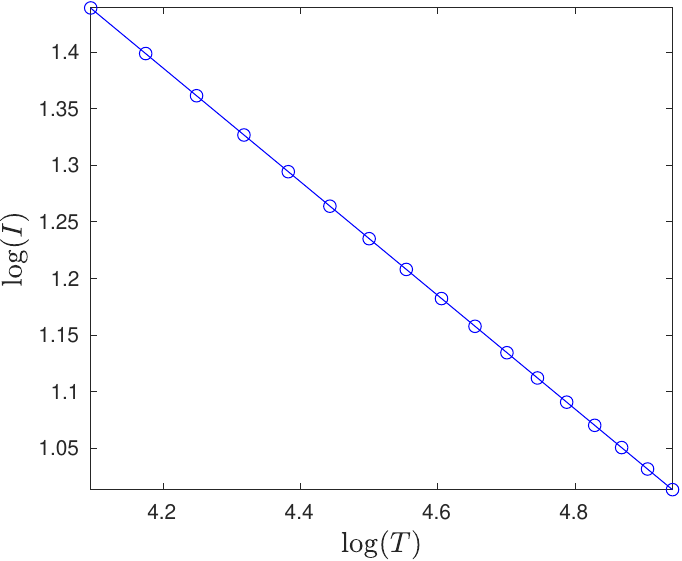}
    \caption{(a)  Log-log plot of the MI $I(A_1, A_2)$ as a function of temperature. The linear fit gives a slope of 0.50(2). The data points are computed at average filling $\nu = 1$, using the high temperature overlap matrix technique on an adjacent geometry where the length of the boundary of subregion $A$ is the same as the boundary of the whole system ($l_y = L_y = L_x = 285$, and $l_x = 10$). }
    \label{fig:mutualInfo_hightemp}
\end{figure}

Fig.\ref{fig:EE_hightemp} shows the thermal entropy computed on subregion $A$ in the adjacent geometry using the high temperature overlap matrix, from which we verify numerically that the leading term indeed obeys the same volume law and scales logarithmically with temperature as Eq.~(\ref{eq:thermalEE}). We also compute the MI to study the behaviour of the high temperature entanglement beyond the volume law. The numerical results at $\nu =1 $ are plotted in Fig.\ref{fig:mutualInfo_hightemp} as a log-log plot. We find that the MI vanishes slowly with respect to temperature, decaying as a power law of temperature with exponent $-0.50(2)$. 

\begin{figure}
    \centering
    \includegraphics[scale=0.55]{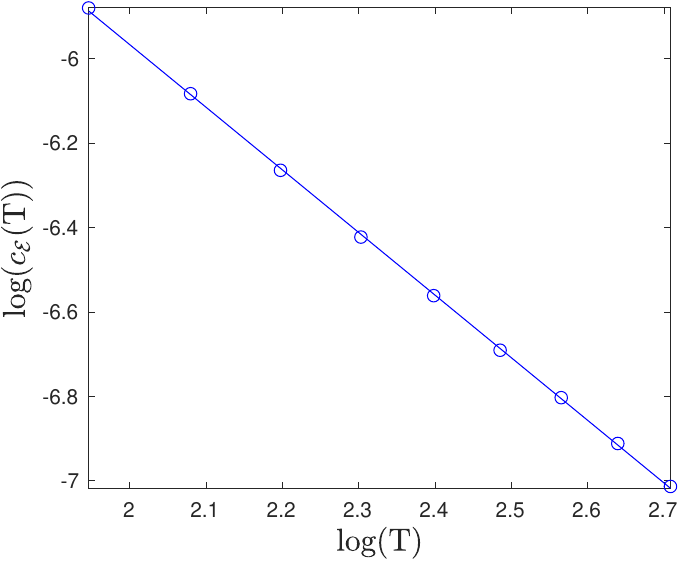}
    \caption{Log-log plot of the high temperature boundary law coefficient $c_\mathcal{E}(T)$ as a function of temperature at average filling $\nu = 1$ on an adjacent geometry with no corners. The boundary law coefficient decays as a power law of temperature, with exponent $-1.5(1)$. The numerical data points are computed from the high temperature overlap matrix technique on an adjacent geometry where the length of the boundary of subregion $A$ is the same as the boundary of the whole system ($l_y = L_x = L_y = 28$ and $l_x = 10$). }
    \label{fig:LN_hightemp}
\end{figure}
Finally, we study the high temperature behaviour of the LN using the high temperature overlap matrix technique on an adjacent geometry of a torus where the length of the subregion $l_y$ equals the length of the torus $L_y$.  Numerically, we observe that the area law (\ref{eq:tempAreaLaw}) persists in this temperature regime. The temperature dependence of the boundary law coefficient, $c_\mathcal{E}(T)$, is plotted in Fig.~\ref{fig:LN_hightemp} as a log-log plot. We find that the boundary law coefficient decays as a power law, $T^{-1.5(1)}$. At high temperatures, we thus see that on the adjacent geometry the LN is smaller than the MI. The emergence of power laws for $T\gg 1$ is natural since the temperature far exceeds the cyclotron energy, so that the electrons behave almost like free particles at finite temperature. The quadratic dispersion leads to a dynamical exponent $z=2$, so that when converting the temperature to a length scale, one has $T^{-1/z}=T^{-1/2}$. It is then natural that the boundary law coefficient scales as a power of this thermal length $T^{-p/2}$, where $p$ is some positive integer. For the MI, we found $p=1$, while for the LN, $p=3$. 
A power law decay is also observed for IQH charge fluctuations where at high temperatures the boundary law vanishes as $T^{-3/2}$ ($p=3$), and the corner term as $T^{-2}$ ($p=4$) as we show in Appendix~\ref{appendix:HighTFluctuations}. 

\section{Discussion}\label{sec:con}
We have studied the non-perturbative properties of the LN (both bosonic and fermionic), and of the MI for isotropic states (mixed or pure) on various geometries with a special focus on the case where two corners touch. The LN and MI were compared with the mutual fluctuations of a local observable, such as the charge. The angle dependence of the three quantities was found to possess similar features such as identical divergences in certain limits, but they yield distinct coefficients that characterize the state. For the MI, some properties were proved generally, owing to SSA, but since the LN is not a convex measure for all density matrices, we also had to rely on heuristic arguments. For instance, we were not able to prove that the LN for the hourglass geometry always increases with the angle. Moreover, the MI and LN on the hourglass were observed to increase in a convex fashion in all cases studied, so it would be interesting to see how general this is. It would be worthwhile to investigate under what conditions such properties can be shown rigorously.  We note that most of the properties are expected to hold for the R\'enyi generalizations of the MI, $I_n(A_1,A_2)=S_n(A_1)+S_n(A_2)-S_n(A_1 A_2)$.

We checked our general results with IQH states at fillings $\nu=1,2$, both a zero and finite temperatures. In the latter case, we found that the gap does protect the MI and LN at asymptotically low $T$, but that they decay fast inside the gap. At large temperatures, we found that the LN decays with a power $T^{-3/2}$, which is the same as for mutual fluctuations,
and is thus parametrically smaller compared with the MI that scales as $T^{-1/2}$. A physical understanding of these power laws would be needed. 

It would be of interest to test our predictions, and to obtain the various coefficients in other states such as in the FQH effect, or other quantum critical systems, including interacting CFTs.
In addition, the overlap matrix method that we developed should be useful to study the LN in other Gaussian states.

\acknowledgements 
 We thank Cl\'ement Berthiere, Rufus Boyack, Gilles Parez and Hassan Shapourian for useful discussions.
 The work is supported by a Discovery Grant from NSERC, a Canada Research Chair, and a grant from the Foundation Courtois. The numerical simulations were enabled in part by support provided by Calcul Québec and Compute Canada.

\newpage
\begin{widetext}

\appendix

\section{Data}\label{appendix:data}
On a smooth bipartite geometry, the eigenvalues of the overlap matrix can be computed analytically. When the subregion is sufficiently large, the eigenvalue sum, Eq.~(\ref{eq:eig_sumEE}) in the main text, can be approximated as an integral, and the boundary law coefficient computed by a numerical integral without being limited by the precision of the numerical diagonalization. See Ref.~\onlinecite{Rodriguez} for details. We compute the boundary law coefficients this way and summarize them in Table~(\ref{Tab:slope}). In this geometry, the LN boundary law coefficient $c_\mathcal{E}$ corresponds to the boundary law coefficient of $S_{1/2}(A_1)$, the Rényi entropy with index 1/2 . 
\begin{table}[!htbp]
\begin{center}
\caption{\label{Tab:slope} Boundary law coefficients for the LN $\mathcal{E}$ and the MI  $I(A_1,A_2)$ in the IQH groundstates at fillings $\nu = 1,2$. Note that these are exactly the R\'enyi-1/2 boundary law coefficients: $c_{\mathcal{E},\nu}=c_{1/2, \nu}$.}
\scalebox{1.2}{
\begin{tabular}{ccc}\toprule[0.5pt]
$c_{\mathcal{E}, \nu = 1}$ & 0.278936335\\
$c_{1, \nu = 1}$ & 0.203290813 \\
$c_{\mathcal{E}, \nu = 2}$&  0.495444054 \\
$c_{1, \nu = 2}$& 0.356989866
\\ \bottomrule[0.5pt]
\end{tabular}}
\end{center}
\end{table}

In the main text, we use two approaches to compute the LN $\mathcal{E}$: an overlap matrix approach and a real space discretization method. The latter method introduces a lattice spacing $a$. The final results thus have to be extracted with a finite-size analysis, by taking $a\rightarrow 0$. Fig.~\ref{fig:negparahourfiniteL10} shows the comparison between the numerical data and the finite-size fitting curve for an example angle, $0.3\pi$, for both the MI and the LN at $\nu = 1$. 


\begin{figure}
    \centering
    \includegraphics[scale=0.55]{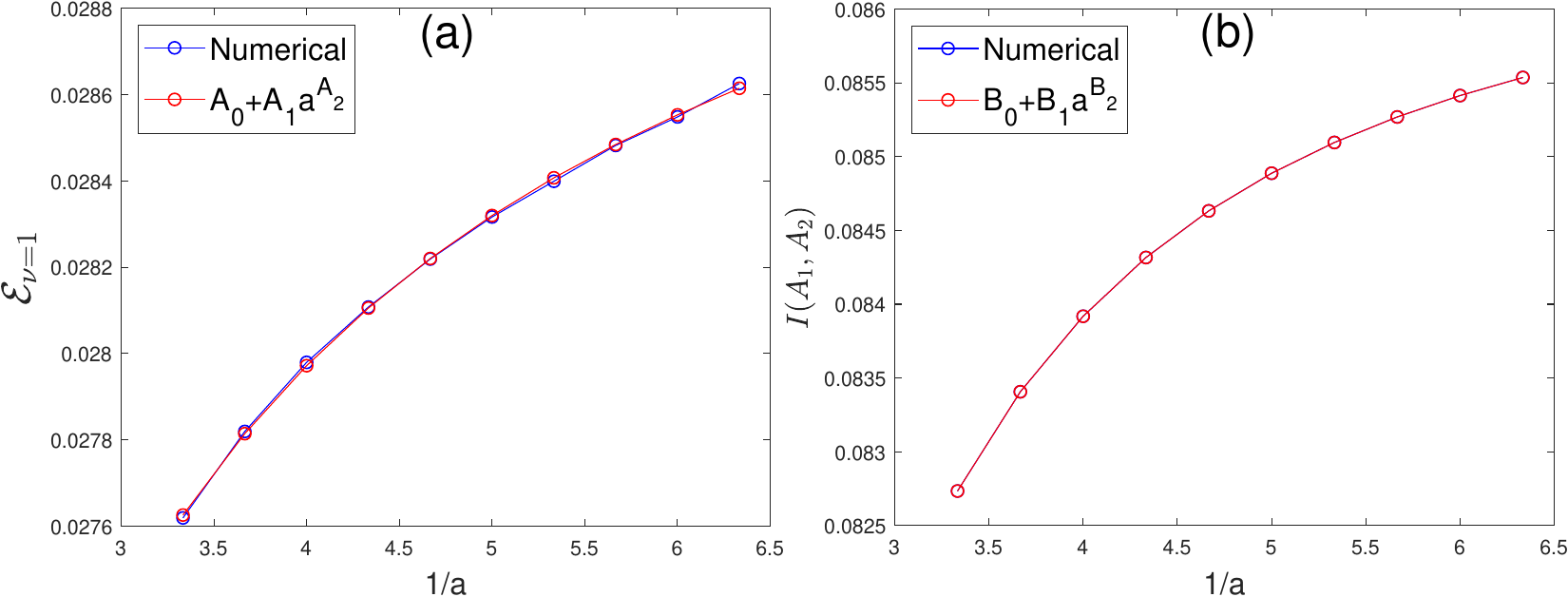}
    \caption{\label{fig:negparahourfiniteL10}The numerical results of (a) the LN $\mathcal{E}$ (b) the MI  $I(A_1,A_2)$ and the corresponding finite-size fits on the hourglass geometry at touching angle $\theta=0.3\pi$ and $\nu=1$ versus the inverse of the lattice spacing $a$, where $A_0=0.030$, $A_1= -0.007$, $A_2=0.973$,  $B_0=0.08661$, $B_1= -04306$, and $B_2=1.99990$. }
\end{figure}

On the other hand, the infinite-dimensional overlap matrices need to be truncated numerically to some size $N_{\text{cutoff}}$, which amounts to ignoring electrons far from the entanglement cut which contribute negligibly to the entanglement. However, one must also be careful about the various length scales in the problem. We must take $L_y$, the size of the torus, to be large enough with respect to the boundary $l_y$ of subregion $A = A_1 A_2$ so that the size of subregion $B$ is much larger than the size of subregion $A$. However, the size of the matrix $N_{\text{cutoff}}$ also grows with the size of subregion $B$ so one must take care not to make $L_y$ too large. One must also be careful that the length of the shared boundary $l_y$ is large enough so that we are in the area law regime (that is, the LN or MI respect the scaling laws Eq.~(\ref{scaling}) and (\ref{MIscaling}), respectively). Thus, one must compute the MI and LN for various $L_y$, $l_y$ and $N_{\text{cutoff}}$ until the numerical results converge to the desired precision. Tables~(\ref{Tab:slope})-(\ref{Tab:hourglass}) show the numerically computed LN $\mathcal{E}$ and MI $I(A_1,A_2)$ data on the adjacent and hourglass geometries, from which we see that both the real space discretization method and the overlap matrix approach give the same results at least up to the third digit. In general, we find it much easier to get the MI to converge than the LN. Nonetheless, at low temperatures and especially at $T = 0$, when the overlap matrices still have manageable numerical sizes, the overlap matrix technique gives more digits of precision than the real space approach for both the LN and the MI.

\begin{figure}
\centering
\includegraphics[scale=0.5]{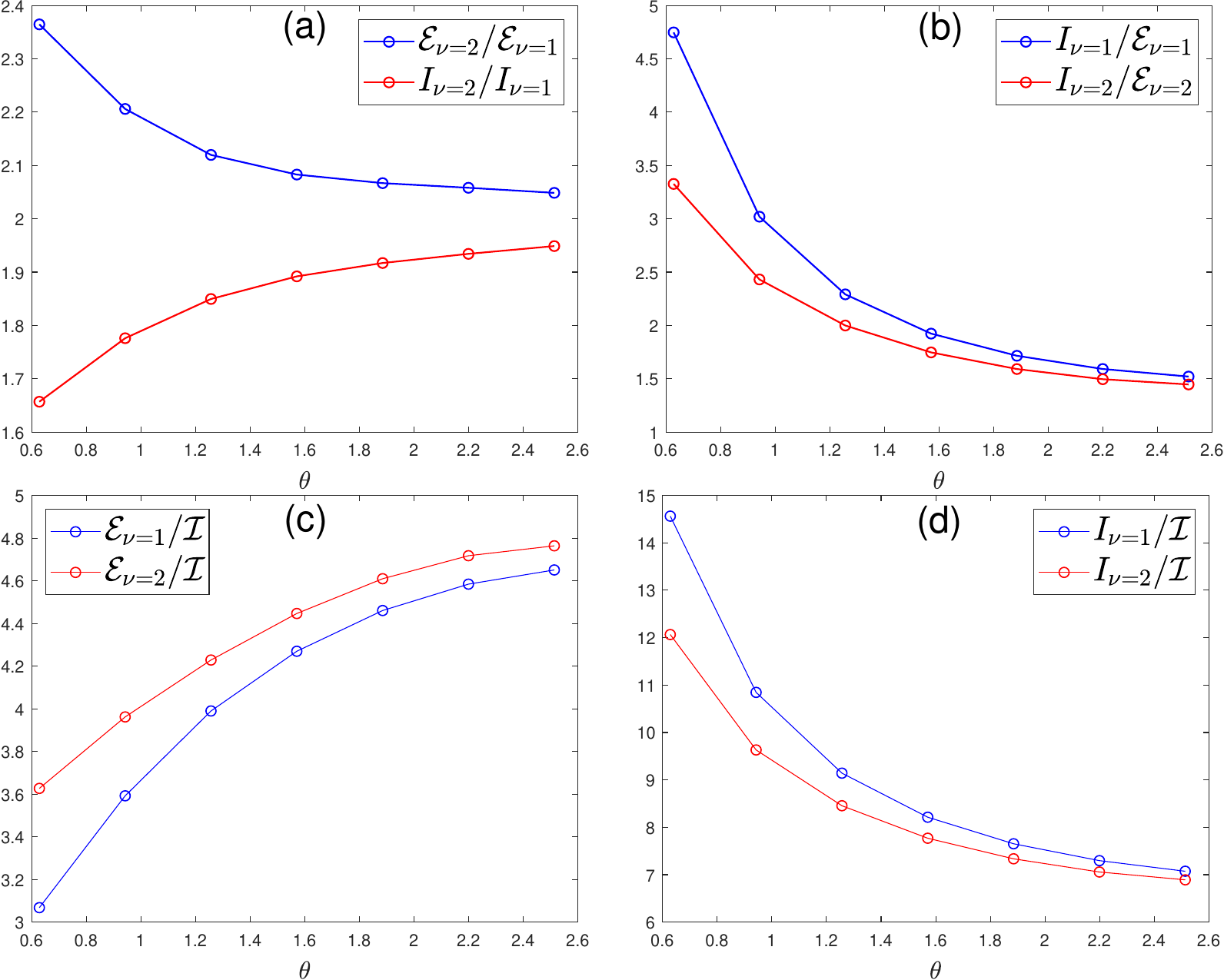}
\caption{\label{fig:ratiohourglass}Various ratios between the LN $\mathcal{E}$, MI $I$, and the mutual charge fluctuations $\mathcal{I}$, on the hourglass geometry as a function of the angle.}
\end{figure}

\red{In Fig.~\ref{fig:ratiohourglass}, we show different ratios for the hourglass geometry. First, in panel (a), we compare the angle dependence between the two fillings. The ratio $\mathcal E_{\nu=2}/\mathcal E_{\nu=1}$ shows a stronger angle dependence compared to what was found for the adjacent geometry, exceeding the naive value 2 by at most $20\%$ (in the range studied). The ratio for the MI behaves like the one for LN but reflected below 2. Panels (b)-(d) show ratios of different quantities at the same filling. We see that the angle dependence shows the most variability at small angles.}


\begin{table}[!htbp]
\begin{center}
\caption{\label{Tab:adjancent}Subleading terms of the MI, $b^I(\theta)$ and the LN, $b(\theta)$ for various angles $\theta$ and fillings $\nu$ on the adjacent geometry, using the overlap matrix and the real space lattice approaches. }
\scalebox{1.2}{
\begin{tabular}{ccccc|cccc}\toprule[0.5pt]
\rule{0pt}{2.5ex} 
   $\theta$  & \multicolumn{4}{c}{$b^I(\theta)$} & \multicolumn{4}{c}{$b(\theta)$}\\
    & \multicolumn{2}{c}{$\nu$ = 1} & \multicolumn{2}{c}{$\nu$ = 2} & \multicolumn{2}{c}{$\nu$ = 1} & \multicolumn{2}{c}{$\nu$ = 2}\\ 
   & Overlap & Lattice & Overlap & Lattice & Overlap & Lattice & Overlap & Lattice\\  \midrule[0.1pt] 
 0.1$\pi$& 0.8492015 & 0.849 & 1.67012 & 1.670 & 0.6618 & 0.661(2)& 1.286 & 1.289(3) \\
 0.2$\pi$& 0.397542 & 0.398 &0.78262 & 0.782 & 0.3139 & 0.314(1)& 0.608 & 0.608(2)   \\
 0.3$\pi$& 0.250490 &  0.251&  0.49354 & 0.493 & 0.2018 & 0.202(1)& 0.389 & 0.390(1)  \\
 0.4$\pi$& 0.188743  &  0.188 & 0.37220 & 0.372 & 0.1551 & 0.155(1)& 0.299 & 0.299(1) \\
 0.5$\pi$& 0.170997 & 0.1710 & 0.33733 & 0.337 & 0.1417 & 0.142(1)& 0.273 & 0.273(1)  \\ \bottomrule[0.5pt]
\end{tabular}}
\end{center}
\end{table}

\begin{table}[!htbp]
\caption{\label{Tab:hourglass}Mutual information $I(A_1,A_2)$ and LN $\mathcal{E}$ for various angles $\theta$, and fillings $\nu$, in the hourglass geometry. Results from the overlap matrix and the real space lattice approaches are shown. }
\begin{center}
\scalebox{1.2}{
\begin{tabular}{ccccc|cccc}\toprule[0.5pt]
\rule{0pt}{2.5ex}
   $\theta$  & \multicolumn{4}{c}{$I(\theta)$} & \multicolumn{4}{c}{$\mathcal{E}(\theta)$}\\
    & \multicolumn{2}{c}{$\nu$ = 1} & \multicolumn{2}{c}{$\nu$ = 2} & \multicolumn{2}{c}{$\nu$ = 1} & \multicolumn{2}{c}{$\nu$ = 2}\\ 
   & Overlap & Lattice & Overlap & Lattice & Overlap & Lattice & Overlap & Lattice\\ \midrule[0.1pt] 
 0.2$\pi$& 0.04987285 &  0.04987  & 0.08264 & 0.08264 & 0.010486 & 0.011(2) & 0.025 & 0.026(2)  \\
 0.3$\pi$&0.0866097 & 0.08661 & 0.15382  & 0.15382 & 0.028663  & 0.030(2)& 0.063 & 0.064(2)   \\
 0.4$\pi$& 0.1370208 & 0.13702 & 0.25344 &  0.25344 & 0.059775 & 0.061(1)& 0.127 & 0.127(1)  \\
 0.5$\pi$& 0.20802678 & 0.20803 &  0.39359 & 0.39359 & 0.108130  &  0.109(1)&  0.225 & 0.225(1)\\ 
  0.6$\pi$& 0.31255950 & 0.31256 & 0.59920 & 0.5992 & 0.182168  & 0.183(1) &0.377 & 0.377(1)   \\
 0.7$\pi$& 0.4801451 & 0.48015 & 0.92872 & 0.9287 & 0.301634 & 0.302(1)& 0.621 & 0.621(1)   \\
 0.8$\pi$& 0.7989219 & 0.7989 & 1.55694 & 1.557 &  0.525306  & 0.526(2) & 1.077 & 1.077(2)  \\ 
 \bottomrule[0.5pt]
\end{tabular}}
\end{center}
\end{table}

\section{Logarithmic negativity for uncorrelated subregions}\label{appendix:LNnocorrelation}

Under the condition that there are no correlations between subregions $A_1$ and $A_2$,
\begin{equation}\label{eq:uncorrelate}
C_{12}=C_{21}\cong 0
\end{equation}
we have
\begin{equation}
\Gamma_{+}=\Gamma_{-}=\begin{bmatrix}
-\Gamma_{11}&\pm i \Gamma_{12}\\
\pm i \Gamma_{21} &\Gamma_{22}
\end{bmatrix}=\begin{pmatrix}
2C_{11}-\mathbb{1}&0\\ 0& \mathbb{1}-2C_{22}
\end{pmatrix},
\end{equation}
which is block-diagonal, and the composite correlation function $C_{\times}$ is
\begin{equation}
C_{\times}=\frac{1}{2}\left(\mathbb{1}-\left(\mathbb{1}+\Gamma_+\Gamma_-\right)^{-1}\left(\Gamma_++\Gamma_-\right)\right)=\frac{1}{2}(\mathbb{1}+\Gamma^2_+)^{-1}(\mathbb{1}-\Gamma_+)^2.
\end{equation}

The composite correlation function $C_{\times}$ and the correlation function $C_{\mathbf{r},\mathbf{r}'}$ can be simultaneously diagonalized, and the eigenvalue $\varepsilon$ of $C_{\times}$ in this case can be expressed in terms of the eigenvalue $\zeta$ of the correlation function $C_{\mathbf{r},\mathbf{r}'}$:
\begin{equation}
\varepsilon_j=\begin{cases}
    \frac{2(1-\zeta_j)^2}{1+(1-2\zeta_j)^2},\quad j\in A_1 \\ \frac{2\zeta_j^2}{1+(1-2\zeta_j)^2},\quad j\in A_2.
  \end{cases}
\end{equation}

As a result, the LN is zero: 
\begin{equation}
\begin{aligned}
&\mathcal{E}=\sum_{j}\log{\left[\varepsilon^{\frac{1}{2}}_j+\left(1-\varepsilon_j\right)^{\frac{1}{2}}\right]}+\frac{1}{2}\sum_j\log{\left[\zeta^2_j+\left(1-\zeta_j\right)^2\right]}= 0 .
\end{aligned}
\end{equation}
 
The condition (\ref{eq:uncorrelate}) is held in the case of two distant subregions at zero temperature, or in the case of two disjoint subregions in high temperature limit, $\beta\rightarrow 0$. In both cases, the LN vanishes.

\section{Low temperature expansion for the overlap matrices}\label{appendix:lowToverlap}

In the low temperature region $T\ll 1$,  the chemical potential for the IQH state at filling $\nu=1$ is almost a constant, $\mu\cong 0.5$, and the Fermi-Dirac distribution $n_F\left(\epsilon_n\right)$ can be expanded in term of $\lambda=e^{-\frac{\beta}{2}}$. Consequently, the leading correction to the correlation function $C_{\mathbf{r},\mathbf{r}'}$ involves only the $0$-th and $1$-st Landau levels:
\begin{equation}
\begin{aligned}
&C_{\mathbf{r},\mathbf{r}'}=\sum_{n,k}n_F\left(\epsilon_n\right)\phi^*_{n,k}(\mathbf{r})\phi_{n,k}(\mathbf{r}')\cong (1-\lambda) C_{0,\mathbf{r},\mathbf{r}'}+\lambda C_{1,\mathbf{r},\mathbf{r}'}\\
&=\left(1-\lambda\right)\sum_k\phi^*_{0,k}(\mathbf{r})\phi_{0,k}(\mathbf{r}')+\lambda\sum_k\phi^*_{1,k}(\mathbf{r})\phi_{1,k}(\mathbf{r}').\\
\end{aligned}
\end{equation}

Following this correlation function, the overlap matrices $F^{(i)}$, where $i=1,2$ corresponds to the overlap matrices on the subregion $A_{1}$ and $A_{2}$  respectively, can be constructed in the space composed of the $0$-th and $1$-st Landau levels:
\begin{equation}\label{eq:overlapLowT}
F^{(i)}=\begin{pmatrix}
(1-\lambda)F^{(i)}_{0,0} & (1-\lambda)F^{(i)}_{0,1} \\
\lambda F^{(i)}_{1,0} & \lambda F^{(i)}_{1,1}
\end{pmatrix},
\end{equation}
where $F^{(i)}_{n,m}$ are block overlap matrices in the $n$-th and $m$-th Landau levels, respectively, with the matrix elements defined as Eq.~(\ref{eq:defoverlap}) in the main text.

\section{Correlation function in high temperature regions}\label{appendix:highT}
We compute the correlation  function $C_{\mathbf{r},\mathbf{r}'}=\sum_{n,k}n_F(\epsilon_n)\phi^*_{n,k}(\mathbf{r})\phi_{n,k}(\mathbf{r}')$ in the high temperature regions $T=1/\beta\gg 1$, where the Fermi-Dirac distribution $n_F(\epsilon)=(1+e^{\beta(n-\mu(\beta))})^{-1}$ can be approximated by the Boltzmann distribution $e^{-\beta(n-\mu(\beta))}$ so that
\begin{equation}\label{eq:corr_sumapprox}
   C_{\mathbf{r},\mathbf{r}'} \approx \frac{e^{\beta \mu(\beta)}}{\sqrt{\pi} L_y}\sum_{k}e^{i k (y' - y)} e^{-\frac{1}{2}(x + k)^2 - \frac{1}{2}(x' + k)^2}\sum_{n=0}^{\infty} \frac{e^{-\beta n}}{2^n n!}  H_n(x + k)H_n(x' + k).
   \end{equation}
We show how to evaluate exactly for the summation over the energy-level $n$  in the high temperature correlation function, Eq.~(\ref{eq:corr_sumapprox}). Consider the following summation,
\begin{equation}
    P = \sum_{n = 0}^{\infty} \frac{e^{-\beta n}}{2^n n!}  H_n(z)H_n(w).
\end{equation}
Writing the Hermite polynomials in the summation using their integral form $  H_n(z) = \frac{(-2i)^n}{\sqrt{\pi}} e^{z^2} \int_{-\infty}^{\infty} u^n e^{-u^2 + 2 i z u } du $, we have

\begin{equation}
\begin{aligned}
    P &= \sum_{n = 0}^{\infty}\frac{e^{-\beta n}}{2^n n!}\Big[\frac{(-2i)^n}{\sqrt{\pi}} e^{z^2} \int_{-\infty}^{\infty} u^n e^{-u^2 +2izu} du \Big] \Big[\frac{(-2i)^n}{\sqrt{\pi}} e^{w^2} \int_{-\infty}^{\infty} v^n e^{-v^2 +2iwv} dv \Big] \\
    & = \frac{1}{\pi} e^{z^2}e^{w^2} \int_{-\infty}^{\infty}dv \int_{-\infty}^{\infty}du e^{-u^2 -v^2 + 2 i z u + 2 i w v } \sum_{n = 0}^{\infty} \frac{e^{-\beta n}}{2^n n!}(-2 i)^{2n}u^n v^n.
\end{aligned}
\end{equation}
Now, the infinite summation over the energy-level $n$   can be done easily, 
\begin{equation}
\sum_{n = 0}^{\infty} \frac{e^{-\beta n}}{2^n n!}(-2 i)^{2n}u^n v^n = e^{-2 e^{-\beta} u v },
\end{equation}
and after the integration we have
\begin{equation}\label{eq:infinitesum}
\begin{aligned}
        P& = \frac{1}{\pi} e^{z^2}e^{w^2} \int_{-\infty}^{\infty}du e^{-u^2 + 2 u i z } \int_{-\infty}^{\infty}dv e^{-v^2 + v(2 i w  - 2 e^{-\beta}u)}\\
       & = \frac{1}{\sqrt{1 - e^{-2\beta}}} e^{-\frac{(z^2 -2 z w e^{\beta} + w^2 )}{ e^{2\beta} - 1}}.
\end{aligned}
\end{equation}

By using the identity (\ref{eq:infinitesum}), one can verify that 
 \begin{equation}\label{eq:corr_sum}
\begin{aligned}
 &    C_{\mathbf{r},\mathbf{r}'} \approx \frac{e^{\beta \mu(\beta)}}{\sqrt{\pi} L_y}\sum_{k}e^{i k (y' - y)} e^{-\frac{1}{2}(x + k)^2 - \frac{1}{2}(x' + k)^2}\sum_{n=0}^{\infty} \frac{e^{-\beta n}}{2^n n!}  H_n(x + k)H_n(x' + k)\\
    &= \frac{e^{\beta \mu(\beta)}}{\sqrt{\pi} L_y}\sum_{k} \Big[e^{i k (y' - y)} e^{-\frac{1}{2}(x + k)^2 - \frac{1}{2}(x' + k)^2}\sqrt{\frac{1}{1 - e^{-2\beta}}}\exp\Big(-\frac{(x + k)^2 -2(x +k) (x' + k) e^{\beta} + (x' + k)^2 }{ e^{2\beta} - 1} \Big) \Big]. 
    \end{aligned}\end{equation}

The correlation function (\ref{eq:corr_sum}) can be  approximated by using the integration to replace the momentum summation in the thermodynamic limit. Moreover, in the high temperature regions $\beta\ll 1$, the chemical potential for the grand canonical ensemble at filling $\nu$ is $\mu(\beta)\approx \frac{1}{\beta}\log{(\nu\beta)}$. As a result, at the filling  $\nu=1$, the correlation function (\ref{eq:corr_sum}) reduces to Eq.~(\ref{eq:highTcorr_limit}).

\section{Overlap matrix method in high temperature regions}\label{appendix:OverlapHighT}

In the limit $\beta\rightarrow 0$, the amplitude of the correlation matrix $C_{\mathbf{r},\mathbf{r}'}$ (\ref{eq:highTcorr_limit}) is centralized near the region $|\mathbf{r}-\mathbf{r}'|^2<2\beta$, where the phase factor is almost vanishing. Therefore, the high temperature correlation function $C_{\mathbf{r},\mathbf{r}'}$ can be further approximated as
\begin{equation}\label{eq:highTcorr_limitAP}
C_{\mathbf{r},\mathbf{r}'} \approx \frac{1}{2\pi}e^{-\frac{1}{2\beta}|\mathbf{r}-\mathbf{r}'|^2}.
\end{equation}
Based on  $C$, Eq.~(\ref{eq:highTcorr_limitAP}), the corresponding composite correlation function $C_{\times}$ in the high temperature region can be constructed by following Eq.~(\ref{eq:associatedC}). The correlation function $C_{\mathbf{r},\mathbf{r}'}$ (\ref{eq:highTcorr_limitAP}) is now separable, which means that
\begin{equation}
C_{\mathbf{r},\mathbf{r}'}=\frac{1}{2\pi}e^{-\frac{1}{2\beta}|\mathbf{r}-\mathbf{r}'|^2}=\sum_{k_x,k_y}\phi^{H, *}_{k_x,k_y}(\mathbf{r})\phi^{H}_{k_x,k_y}(\mathbf{r}'),
\end{equation}
where
\begin{equation}
\phi^{H}_{k_x,k_y}(\mathbf{r})=\sqrt{\frac{\beta}{L_xL_y}} e^{-\frac{\beta\left(k_x^2+k_y^2\right)}{4}-ik_xx-ik_yy}=f_{k_x}(x)f_{k_y}(y),
\end{equation}
and
\begin{equation}
f_{k_{x}}(x)=\frac{\beta^{\frac{1}{4}}}{\sqrt{L_{x}}}e^{-\frac{\beta k_x^2}{4}-ik_{x}x}, \qquad
f_{k_{y}}(y)=\frac{\beta^{\frac{1}{4}}}{\sqrt{L_{y}}}e^{-\frac{\beta k_y^2}{4}-ik_{y}y}.
\end{equation}
As a result, the spectrum $\lbrace\zeta^H\rbrace$ of $C_{\mathbf{r},\mathbf{r}'}$  can be computed through the overlap matrix:
\begin{equation}\label{eq:OverlapHT}
F^H_{( k_x,k_y),( k'_x,k'_y)}=\int_A d^2\mathbf{r}\phi^{H,*}_{k_x,k_y}(\mathbf{r})\phi^{H}_{k'_x,k'_y}(\mathbf{r}'),
\end{equation}
and the spectrum $\lbrace\varepsilon^H\rbrace$ of $C_{\times}$ can also be extracted through the overlap matrices by following the methodology in Sec.  (\ref{sec:overlap}).

On a rectangle where the two-dimensional integration can be decomposed as two independent one-dimensional integrations, the overlap matrix $F^H_{( k_x,k_y),( k'_x,k'_y)}$ can be further written as a tensor product,
\begin{equation}
F^H_{( k_x,k_y),( k'_x,k'_y)}=F^H_{k_x,k'_x}\otimes F^H_{k_y,k'_y}
\end{equation}
where
\begin{equation}
F^H_{k_x,k'_x}=\int dxf^*_{k_{x}}(x)f_{k'_{x}}(x), \qquad F^H_{k_y,k'_y}=\int dyf^*_{k_{y}}(y)f_{k'_{y}}(y).
\end{equation}
Therefore, in this case, the eigenvalue $\zeta^H_{k_x,k_y}$ is just a product
\begin{equation}
\zeta^H_{k_x,k_y}=\zeta^H_{k_x}\zeta^H_{k_y}
\end{equation}
among the eigenvalues $\zeta^H_{k_{x}}$ and $\zeta^H_{k_y}$ of the overlap matrices $F_{k_x,k'_x}$ and $F_{k_y,k'_y}$, respectively.\\

Based on the overlap matrix (\ref{eq:OverlapHT}), we can derive the spectrum $\lbrace\zeta^H\rbrace$ of $C_{\mathbf{r},\mathbf{r}'}$ analytically on a torus $D$:\begin{equation}
D: \: 0\leq x \leq L_x,\: 0\leq y\leq L_y.
\end{equation}
To satisfy the boundary condition on the torus $D$, we must have
\begin{equation}
k_x=\frac{2\pi n}{L_x},\qquad k_y=\frac{2\pi m}{L_y}
\end{equation}
where $n,m$ are integers. The overlap matrix is thus diagonal
\begin{equation}
F^H_{( k_x,k_y),( k'_x,k'_y)}=\delta_{k_x,k'_x} \delta_{k_y,k'_y}\zeta^H_{k_x,k_y}
\end{equation}
with the eigenvalue
\begin{equation}
\zeta^H_{k_x,k_y}=\beta e^{-\frac{\beta( k_x^2+k_y^2)}{2}}.
\end{equation}
As a result, in the high temperature limit $\beta\rightarrow 0$, the thermal entropy $S_{A}(T)$ on a torus is
\begin{equation}
\begin{aligned}
&S_{A}(T)=\sum_{k_x,k_y}H(\zeta^H_{k_x,k_y})=\frac{L_yL_x}{4\pi^2}\int^{\infty}_{-\infty} dk_x \int^{\infty}_{-\infty}dk_y H\left(\beta e^{-\frac{\beta( k_x^2+k_y^2)}{2}}\right)\\
&\cong -\frac{L_yL_x}{4\pi^2}\int^{\infty}_{-\infty} dk_x \int^{\infty}_{-\infty}dk_y  e^{-\frac{\beta( k_x^2+k_y^2)}{2}} \beta\log{\beta}= -\frac{L_yL_x}{4\pi^2}\frac{2\pi}{\beta}\beta\log{\beta}=\frac{|A|}{2\pi}\log{T},
\end{aligned}
\end{equation}
where $|A| = L_x L_y$ and $H(x)=-x\log{x}-(1-x)\log{(1-x)}$.

\section{Temperature dependence of IQH charge fluctuation boundary law and corner coefficient}\label{appendix:HighTFluctuations}

The charge fluctuations, $\mathcal{F}(A)$, of a system in some subregion $A$ generally scale as \cite{Estienne,Song}
\begin{equation}
      \mathcal{F}(A) = \alpha_{0, \rm fluc}|A| + c_{\rm fluc}|\partial A| - \alpha\, \left(1+ (\pi-\theta)\cot\theta \right)+ \cdots
\end{equation}
The first term scales with the volume $A$, the second term scales with the boundary of $A$ and the third term is the universal corner function, $a_{\rm fluc}(\theta)$, Eq.~(\ref{eq:a-fluc}). In this appendix, we study the temperature dependence of the charge fluctuation boundary law coefficient, $c_{\rm fluc}$, and corner coefficient, $\alpha$, of $\nu = 1$ IQH states.
In particular, we show that at high temperatures, the boundary law coefficient decays as $T^{-3/2}$ while the corner coefficient decays as $T^{-2}$. The temperature dependence of the charge fluctuations volume law coefficient and corner coefficient have previously been studied in Ref.~\onlinecite{Estienne}. 

The boundary law coefficient and corner coefficient  of fluctuations are given by radial integrals of $f(r)$, the connected two-point correlation function which depends on the system under consideration. Specifically, in two dimensions, the boundary law and corner coefficient are given by \cite{Estienne}
\begin{align}
\label{eq:cfluc}
    c_{\rm fluc} &= -2\int_0^\infty \ dr \ r^2 f(r),\\
\label{eq:bfluc}
    \alpha &= -\frac{1}{2}\int_0^\infty \ dr \ r^3 f(r).
\end{align}

For $\nu = 1 $ IQH state at finite temperature, the connected two-point correlation function is (in units of magnetic length $l_B = 1$)
\begin{equation}
\label{eq:finiteT_fr}
f(r, T) = \frac{1}{2\pi}\sum_{n = 0}^\infty n_F(\epsilon_n) \delta(\mathbf{r}) - \frac{e^{-\frac{r^2}{2}}}{4\pi^2}\sum_{n,k = 0}^\infty n_F(\epsilon_n) L_n^0\left(\frac{r^2}{2}\right)n_F(\epsilon_k)L_k^0\left(\frac{r^2}{2}\right),
\end{equation}
where $L_{n}^{0}$ is the associated Laguerre polynomial, and $n_F(\epsilon_n)$, the Fermi-Dirac distribution, captures the temperature dependence. The integral Eq.~(\ref{eq:bfluc}) can be solved analytically for the corner coefficient
\begin{equation}
\label{eq:btemp}
    \alpha(T) =\frac{1}{4\pi^2} \sum_{n = 0}^\infty \left[ n_F(\epsilon_n)^2(2n + 1) - n_F(\epsilon_n)f(\epsilon_{n + 1})(n + 1)\right] .
\end{equation}

In the high temperature limit, we directly work with the high temperature correlation function, Eq.~(\ref{eq:highTcorr_limit}) in the main text, and compute the high temperature charge density correlator from Wick's theorem:
\begin{equation}
    f(\mathbf{r}, T\rightarrow{\infty})\approx \frac{1}{2\pi}\delta(\mathbf{r})  - \frac{1}{4\pi^2}e^{-\frac{r^2}{\beta}}.
\end{equation}

We can then evaluate the integrals in Eq.~(\ref{eq:cfluc}) and Eq.~(\ref{eq:bfluc}). The contribution from the first term is vanishing and we obtain, 
\begin{align}
\label{eq:highTcfluc}
    c_{\rm fluc}\left(T\rightarrow{}\infty\right) &= \frac{1}{8\pi^{3/2}}T^{-3/2},\\
\alpha \left(T\rightarrow{}\infty\right) &= \frac{1}{16\pi^{2}}T^{-2}.\label{eq:highTbfluc}
\end{align}

We show the results in Fig.~\ref{fig:highT_fluc} as log-log plots. For the finite temperature boundary law term, $c_{\rm fluc}(T)$, we numerically evaluate the integral in Eq.~(\ref{eq:cfluc}). Like in the main text, we work in the grand canonical ensemble, with the chemical potential solved self-consistently for a given temperature, $T$. We also plot the results for the temperature dependence of the corner term, $\alpha(T)$, Eq.~(\ref{eq:btemp}). For both the boundary and corner coefficient, the power laws Eq.~(\ref{eq:highTcfluc}) and Eq.~(\ref{eq:highTbfluc}) are in good agreement with the high temperature numerical data.  
\begin{figure}[H]
    \centering
    \includegraphics[scale=0.5]{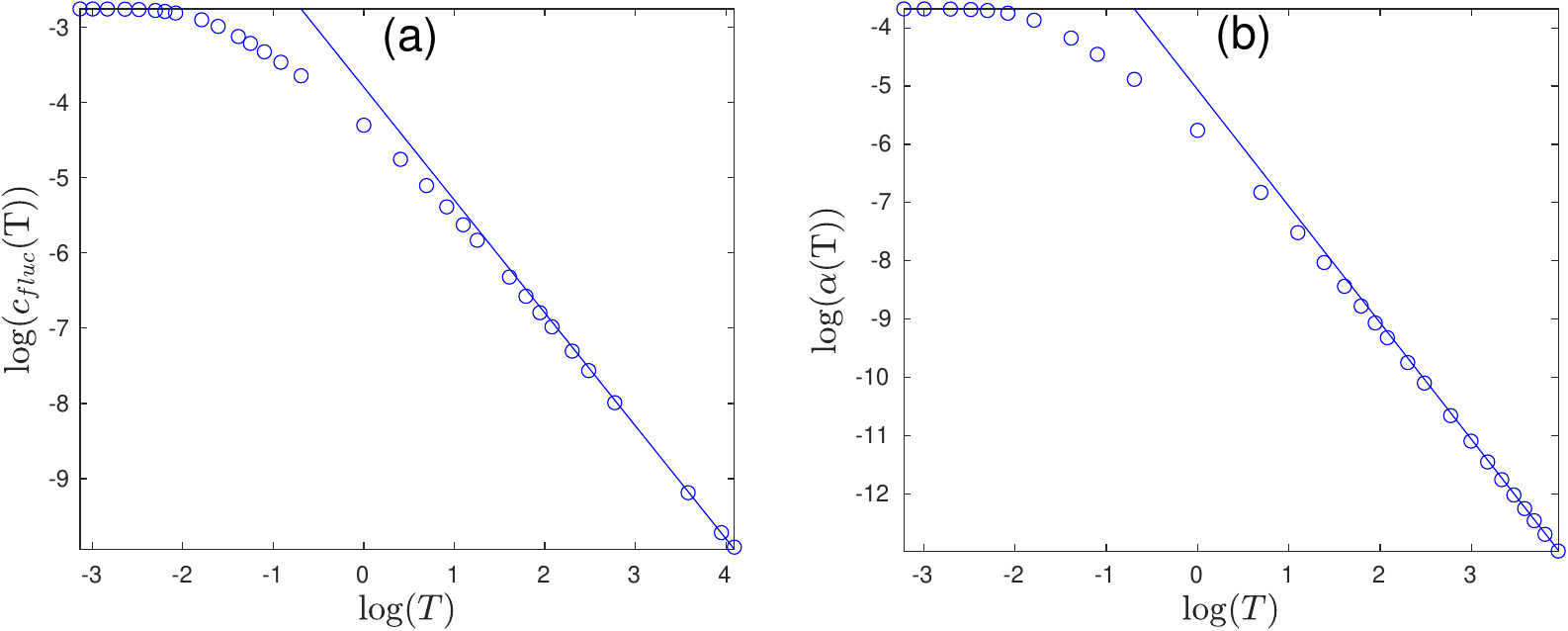}
    \caption{\label{fig:highT_fluc}
Log-log plots of (a) the boundary law coefficient, $c_{\rm fluc}(T)$ and (b) the corner coefficient, $\alpha(T)$ as a function of temperature for IQH charge fluctuations at average filling $\nu = 1$. The solid lines represent the high temperature power laws, Eq.~(\ref{eq:highTcfluc}) and (\ref{eq:highTbfluc}), which are proportional to $T^{-3/2}$  for the boundary law coefficient and $T^{-2}$ for the corner coefficient. }
\end{figure}

\end{widetext}

\bibliography{LNBib}

\begin{thebibliography}{52}
\bibitem [{foo()}]{footnote1}%
  \BibitemOpen
  \href@noop {} {\bibinfo  {journal} {This is true even for many-body pure
  states. One can verify that by expressing the fermionic PT in the fermionic
  occupation number basis. See Ref. \onlinecite{Shapourian3}}\ }\BibitemShut
  {NoStop}%
\end{thebibliography}

\end{document}